%% file: main.tex
\documentclass[11pt]{article}
\usepackage[utf8]{inputenc}
\usepackage{graphicx}
\usepackage{amsmath}
\usepackage{amssymb}
\usepackage{tabularx}
\usepackage[dvipsnames]{xcolor}
\usepackage{mwe}
\usepackage{wrapfig}
\usepackage{multirow}
\usepackage{float}
\usepackage[hyphens]{url}
\usepackage{hyperref}

\RequirePackage[left=1in,%
                right=1in,%
                top=1in,%
                bottom=1in,%
                headheight=12pt,%
                letterpaper]{geometry}%
\usepackage{enumitem}
\usepackage{booktabs}
\usepackage[most]{tcolorbox}
\usepackage{caption}
\usepackage{url}
\title{\Large\bfseries Tutorial: Grid-Following Inverter for Electrical Power Grid}
\author{
\large
Muhammad Hamza Ali, Amritanshu Pandey\\[0.5ex]
\normalsize
Department of Electrical and Biomedical Engineering  \\
University of Vermont \\
Burlington, Vermont
}
\date{}
\begin{document}

\maketitle

\section*{Abstract}
The growing use of inverter-based resources in modern power systems has made grid-following inverters a central topic in power-system modeling, control, and simulation. Despite their widespread deployment, introductory material that explains grid-following inverter operation from first principles and connects control design to time-domain simulation remains limited. To address this need, this tutorial presents a circuit-theoretic introduction to the modeling and simulation of a grid-following inverter connected to an electrical power grid. We describe the inverter’s synchronization with the grid (PLL), power control, and current control structure and show how these elements can be represented within an electromagnetic transient (EMT) simulation framework using companion-model-based formulations similar to those used in circuit simulators such as SPICE and Cadence. We previously wrote a similar tutorial for a grid-connected induction motor \cite{pandey2023circuit}. In this tutorial, we use the grid-following inverter as the primary example to illustrate how its governing equations, control loops, and network interface can be formulated and simulated from first principles. By the end of the document, readers should gain a clear introductory understanding of how to model and simulate a grid-following inverter in an EMT platform.

\section*{Grid-Connected Inverter}
\begin{figure}[H]
    \centering
    \includegraphics[width=0.8\textwidth]{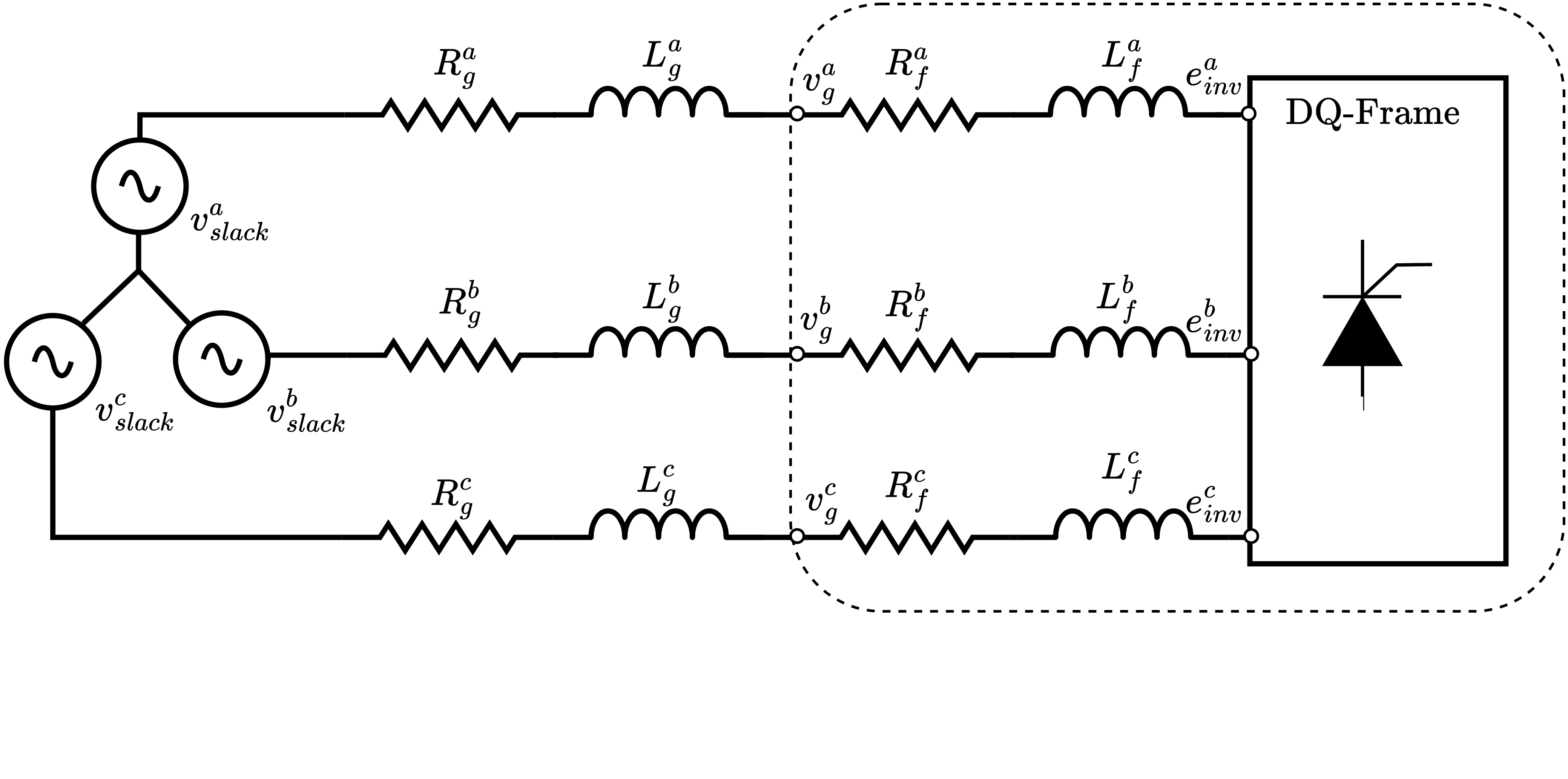} 
    \caption{Grid-connected inverter layout.}
    \label{fig1:inverter_diagram}
\end{figure}

In this study, we interface the grid-connected inverter with an ideal slack-voltage source via a network impedance. 
We model the network impedance with a series resistance $R_g$ and inductance $L_g$, which represent the aggregated transmission or feeder impedance at the point of common coupling (PCC). 
On the converter side, we use an $R_f$--$L_f$ filter to filter the inverter output current and attenuate high-frequency components. Figure~\ref{fig1:inverter_diagram} illustrates the overall configuration.

\noindent This class of converter is generally referred to as a voltage-source converter (VSC) because its control structure is designed to emulate a voltage source within the power system. A VSC may be represented using either a detailed switching model, which explicitly captures the on/off behavior of the semiconductor devices, or an averaged model, which replaces that fast switching action with its averaged effect over each switching interval. In this tutorial, we adopt the averaged model representation, a well-established assumption in power electronics modeling \cite{erickson2007fundamentals}. This approach is appropriate because our primary objective is to study the grid-following inverter dynamics, control behavior, and interaction with the surrounding network, rather than the internal high-frequency switching transients.

\begin{figure}[H]
    \centering
    \includegraphics[width=0.8\textwidth]{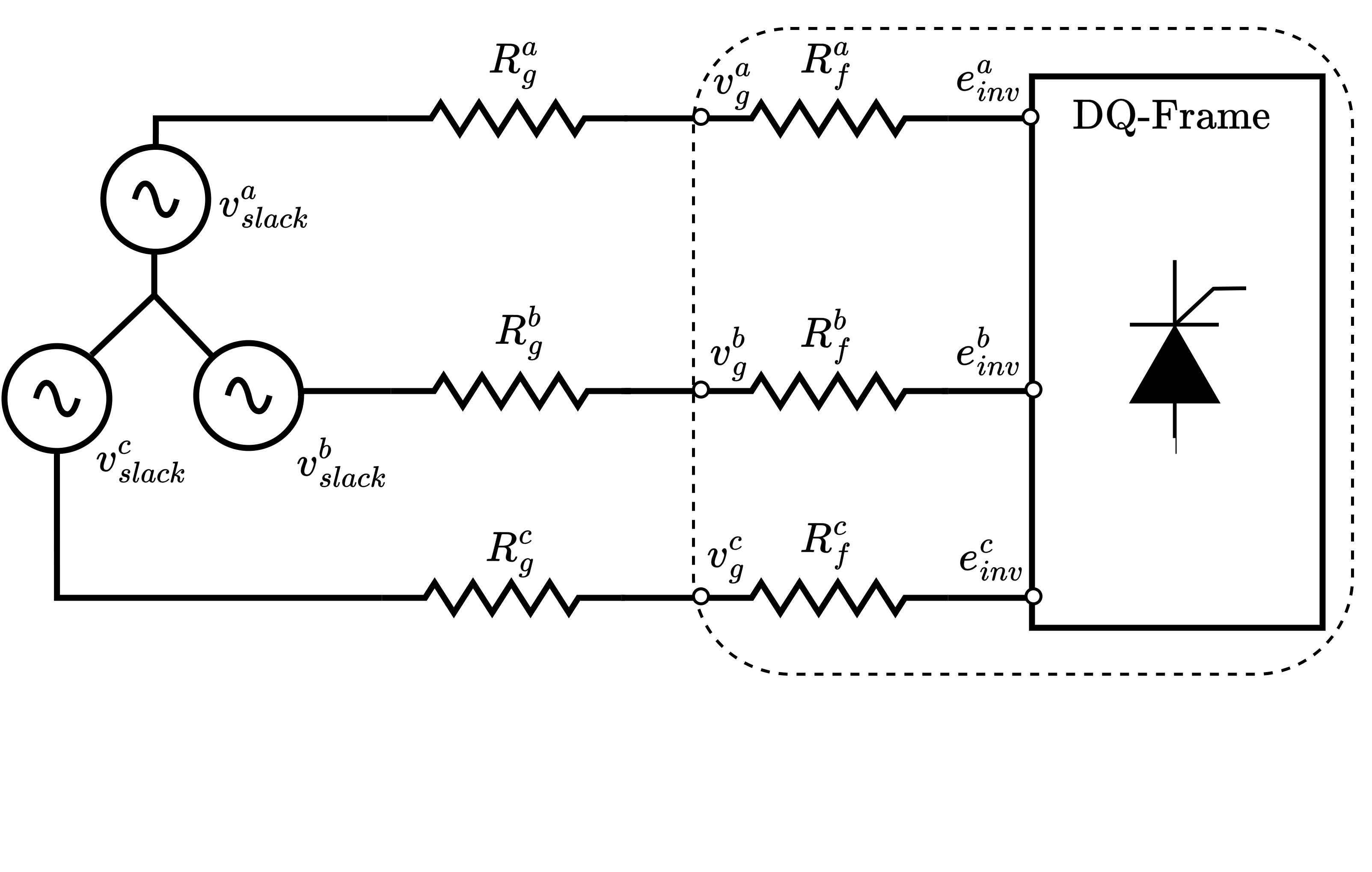} 
    \caption{Grid-connected inverter layout with a resistive element (\textit{Simplified Design}).}
    \label{fig2:inverter_diagram_R}
\end{figure}

Two primary operating modes are distinguished in inverter-based resources (IBRs): grid-following, which synchronizes to an existing voltage source and requires it to deliver controlled active and reactive power; and grid-forming, which behaves as a voltage source that establishes the system’s voltage magnitude, phase angle, and frequency. Because grid-following units cannot operate stably without an external reference, a system composed entirely of grid-following IBRs would require a synchronization signal. Modern power systems with high IBR penetration, therefore, rely on a combination of grid-following and grid-forming inverters to maintain stable operation. 
To understand the inverter grid interconnection, we first analyze a simplified circuit consisting of purely resistive elements in the network and filter, as shown in Fig.~\ref{fig2:inverter_diagram_R}. 

The point $v_{g}^{abc}$ is defined as the point of common coupling, where the inverter injects or draws power to/from the grid. 
The expression for calculating it $v_g^{abc}$ can be derived by applying the KCL at PCC. 
\begin{subequations}
    \begin{align}
        v_g^a (t) &= \cfrac{\big(e_{inv}^a(t) R_g^a + v_{slack}^a(t) R_f^a\big)}{(R_g^a + R_f^a)}  \\
        v_g^b (t) &= \cfrac{\big(e_{inv}^b(t) R_g^b + v_{slack}^b(t) R_f^b\big)}{(R_g^b + R_f^b)}  \\
        v_g^c (t) &= \cfrac{\big(e_{inv}^c(t) R_g^c + v_{slack}^c(t) R_f^c\big)}{(R_g^c + R_f^c)}  
    \end{align}
\end{subequations}

The current leaving the inverter terminal $e_{inv}$ can be found by applying Ohm's law:
\begin{subequations}\label{eq:igdq_res}
    \begin{align}
        i_{g}^a(t) = \cfrac{e_{inv}^a(t) - v_{slack}^a(t)}{(R_g^a + R_f^a)} \\
        i_{g}^b(t) = \cfrac{e_{inv}^b(t) - v_{slack}^b(t)}{(R_g^b + R_f^b)} \\
        i_{g}^c(t) = \cfrac{e_{inv}^c(t) - v_{slack}^c(t)}{(R_g^c + R_f^c)} 
    \end{align}
\end{subequations}

To understand how the inverter injects power into the grid, we examine the internal control structure of the grid-following inverter.

\input{Section/Grid_following_control}
\input{Section/Park_tranformation}

\input{Section/PLL}
\input{Section/Power_controller}
\input{Section/Current_controller_new}

\input{Section/Grid_support}
\newpage
\bibliographystyle{IEEEtran}
\bibliography{biblography}
\newpage
\input{Section/Initial_conditions}

\end{document}

%% file: Section/Grid_following_control.tex
\section*{Grid Following Control}
\begin{figure}[H]
    \centering
    \includegraphics[width=0.8\textwidth]{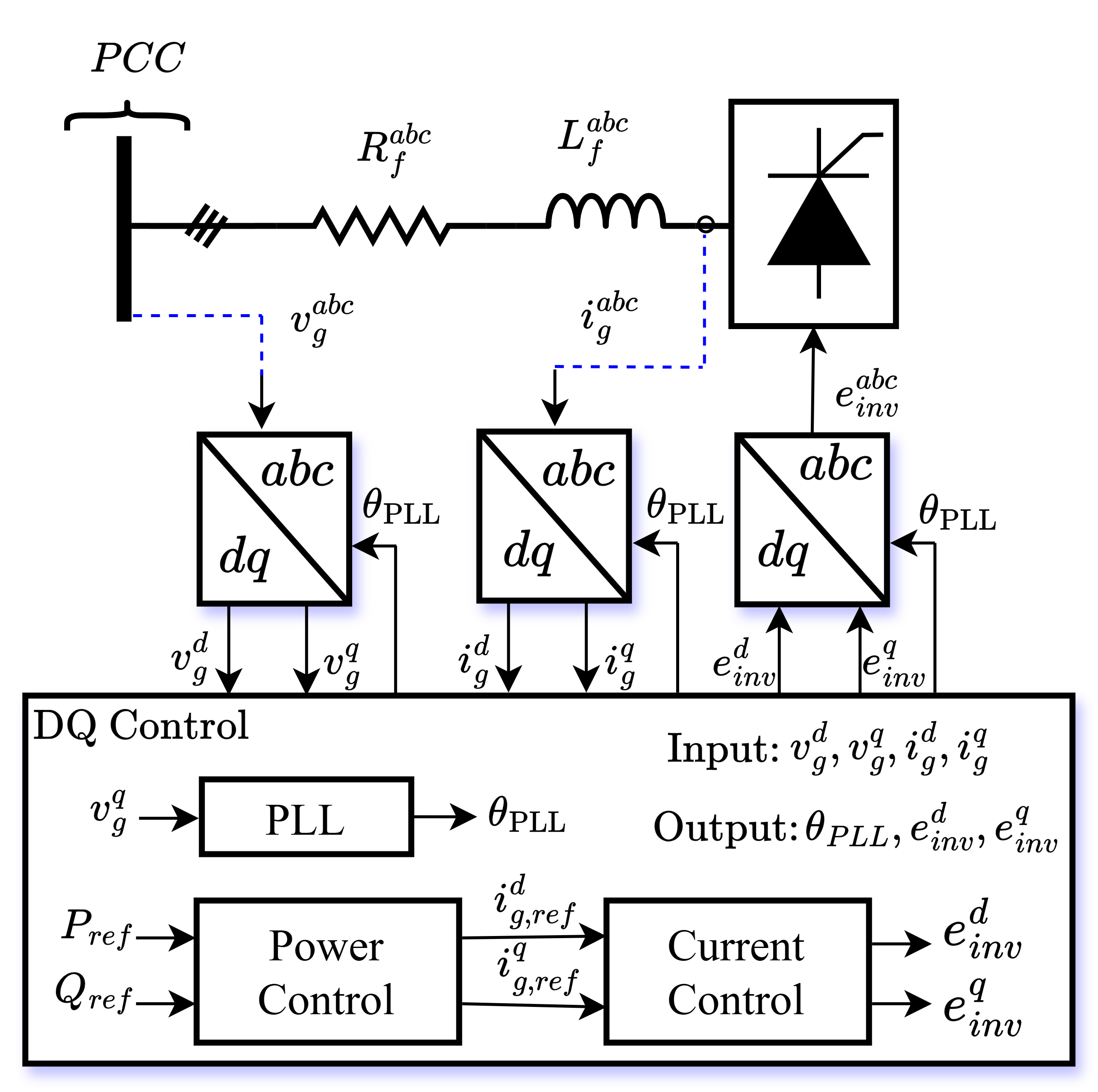} 
    \caption{Grid-connected inverter control diagram layout.}
    \label{fig2:general_diagram}
\end{figure}

The general control structure of the applied grid-following inverter control is shown in Fig.\ref{fig2:general_diagram}. The output of the controller is the three-phase converter voltage $e_{AC}^{abc}$, The grid-current control loop constitutes the lower-level controller and adjusts the internal voltage reference $e_{AC}^{dq}$ such that the grid-current reference $i_{ref}^{dq}$ is accurately tracked. The upper-level power control determines the current references based on the desired active and reactive power injections $P_{ref}$, $Q_{ref}$, while synchronization with the grid is ensured through the phase-locked loop (PLL) using the transformed grid voltages $v_{g}^{dq}$.  

The block descriptions are the following:
\begin{enumerate}[label=\textbf{\arabic*-}]
\item \textbf{Park Transformation} \\
Converts the three-phase grid voltages or currents from the $abc$
frame into the rotating dq reference frame\footnote{direct-quadrature-zero transformation: \url{https://en.wikipedia.org/wiki/Direct-quadrature-zero_transformation}} and vice versa. This transformation enables the controller to operate on DC quantities in the synchronous frame, simplifying the control of active and reactive power.

\item \textbf{Phase-Locked Loop (PLL)} \\
The goal of PLL is to extract the phase from the voltage signal, necessary for the Park Transformation, and allow the controller to independently regulate active and reactive power relative to the grid.

\item \textbf{Power Control} \\
Serves as the upper-level control layer, generating the current references $I_{ref}^{dq}$ based on the specified active and reactive power commands $P_{ref}$, $Q_{ref}$. This block ensures that the desired power is delivered to the grid at the point of common coupling (PCC).
\item \textbf{Grid-Current Control} \\
It operates as the lower-level controller, producing the internal voltage reference $e_{AC}^{dq}$, It regulates the injected currents by adjusting the converter’s terminal voltage so that the grid current follows the reference values determined by the power control block.

\end{enumerate}

%% file: Section/Park_tranformation.tex
\section*{Park Transformation}
This section first defines the three-phase grid voltages that serve as input to the Park transformation. 
Let $\mathbf{v}_g^{abc}(t) = \big[v_g^a(t)\; v_g^b(t)\; v_g^c(t)\big]^\top$ denote the three-phase bus voltages at the node to which the inverter is connected. 
Assuming a nominal frequency of $f = 60~\text{Hz}$, the corresponding angular frequency is $\omega$, and the grid angle $\theta_g(t)$ is expressed in radians. 
The quantity $V_m$ denotes the peak phase voltage in per-unit. 
The grid voltages are modeled as a balanced set of sinusoidal waveforms:
\begin{subequations}\label{eq:grid_voltage}
\begin{align}
    \omega &= 2\pi f, \label{eq:grid_voltage_a} \\
    \theta_g(t) &= \omega t + \theta_{\text{grid}}(t), \label{eq:grid_voltage_b} \\
    v_g^{a}(t) &= V_m \cos\!\big(\theta_g(t)\big), \label{eq:grid_voltage_c} \\
    v_g^{b}(t) &= V_m \cos\!\big(\theta_g(t) - \tfrac{2\pi}{3}\big), \label{eq:grid_voltage_d} \\
    v_g^{c}(t) &= V_m \cos\!\big(\theta_g(t) + \tfrac{2\pi}{3}\big), \label{eq:grid_voltage_e}
\end{align}
\end{subequations}
where $\theta_{\text{grid}}(t)$ represents the angle in radians of the bus to which the inverter is connected.
\begin{tcolorbox}[
    colback=gray!10,
    colframe=gray!50,
    title=Grid Angle Perturbation for PLL Testing,
    title style={color=black},
    enhanced
]
In later simulations, the grid angle $\theta_{\text{grid}}(t)$ is intentionally perturbed to evaluate the dynamic performance of the PLL. A phase jump of magnitude $\Delta\theta_{\mathrm{dist}}$ is introduced at disturbance time $t_{\mathrm{dist}}$ resulting in the following equation:
\[
\theta_{\text{grid}}(t) =
\begin{cases}
0, & t < t_{\mathrm{dist}}, \\[2mm]
\Delta\theta_{\mathrm{dist}}, & t \ge t_{\mathrm{pj}} ,
\end{cases}
\]
\end{tcolorbox}

\begin{figure}[H]
    \centering
    \includegraphics[width=0.8\textwidth]{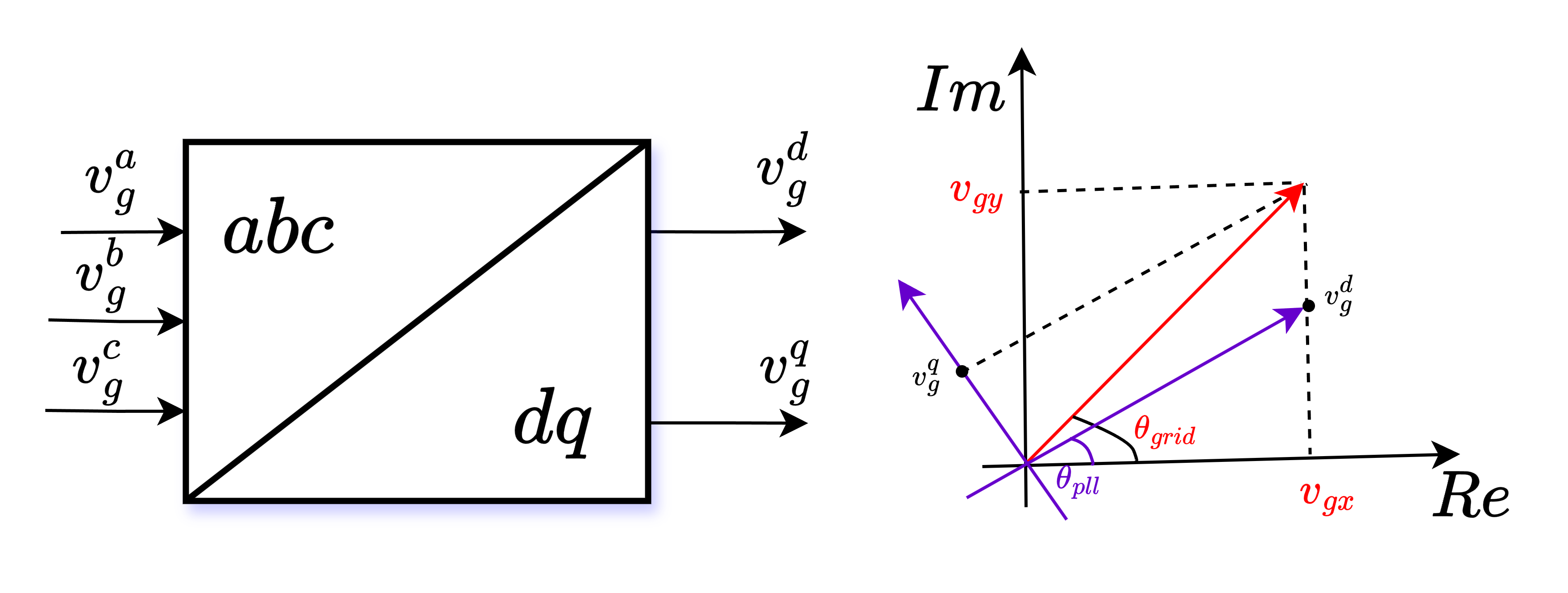} 
    \caption{Park transformation block diagram.}
    \label{fig3:Park transformation}
\end{figure}

Figure~\ref{fig3:Park transformation} illustrates the Park transformation applied to the three-phase
grid voltages $\mathbf{v}_g^{abc}(t) = [v_g^{a}(t)\;\; v_g^{b}(t)\;\; v_g^{c}(t)]^\top$.

\noindent The phasor diagram on the right shows the grid-voltage vector in the complex plane: the red vector represents the actual grid voltage at angle $\theta_{\text{grid}}(t)$,
while the purple axes denote the rotating $dq$ frame aligned with the PLL angle estimate $\theta_{\mathrm{PLL}}(t)$. The values $v_g^{d}$ and $v_g^{q}$ correspond to the
projections of the grid-voltage vector onto these rotating axes.

\medskip

\noindent The PLL constructs an estimated grid angle $\hat{\theta}_g(t)$, which is later used as the
reference for the Park transformation. The PLL angle is defined as
\begin{equation}\label{eq:pll_angle}
    \hat{\theta}_{pll}(t) = \omega t + \theta_{\mathrm{PLL}}(t),
\end{equation}
where $\omega = 2\pi f$ is the nominal angular frequency and $\theta_{\text{pll}}(t)$ denotes the PLL-estimated phase deviation.
Using $\hat{\theta}_g(t)$, the synchronous-frame voltages $v_g^{d}(t)$ and $v_g^{q}(t)$
can be obtained from the three-phase voltages via the Park transformation:
\begin{subequations}\label{eq:park_pll}
\begin{align}
    v_g^{d}(t) &= \frac{2}{3}\Big[
        v_g^{a}(t)\cos\!\big(\hat{\theta}_{pll}(t)\big)
      + v_g^{b}(t)\cos\!\big(\hat{\theta}_{pll}(t) - \tfrac{2\pi}{3}\big)
      + v_g^{c}(t)\cos\!\big(\hat{\theta}_{pll}(t) + \tfrac{2\pi}{3}\big)
    \Big], \label{eq:park_pll_d} \\[2mm]
    v_g^{q}(t) &= -\frac{2}{3}\Big[
        v_g^{a}(t)\sin\!\big(\hat{\theta}_{pll}(t)\big)
      + v_g^{b}(t)\sin\!\big(\hat{\theta}_{pll}(t) - \tfrac{2\pi}{3}\big)
      + v_g^{c}(t)\sin\!\big(\hat{\theta}_{pll}(t) + \tfrac{2\pi}{3}\big)
    \Big]. \label{eq:park_pll_q}
\end{align}
\end{subequations}

%% file: Section/PLL.tex
\section*{Phase locked loop (PLL)}
The phase-locked loop (PLL) provides the angle required for the Park transformation. In grid-following control (\textit{generally}\footnote{It's also possible to set the d-axis to zero, but it requires the reformulation of power/current controller equations}), the $d$ axis of the synchronous $dq$ frame is aligned with the grid-voltage phasor at the PCC, meaning that the Park transformation angle coincides with the actual grid angle. Under this alignment, active-power control is effectively achieved through manipulation of the d-axis current component. In phasor-domain simulations, the PLL is sometimes omitted because the system voltage angle is directly available from the solver. However, during disturbed or unbalanced operating conditions, particularly in systems with high IBR penetration, the PLL dynamics may become nonlinear or even unstable, making explicit PLL modeling essential for stability assessment.

\begin{figure}[H]
    \centering
    \includegraphics[width=0.8\textwidth]{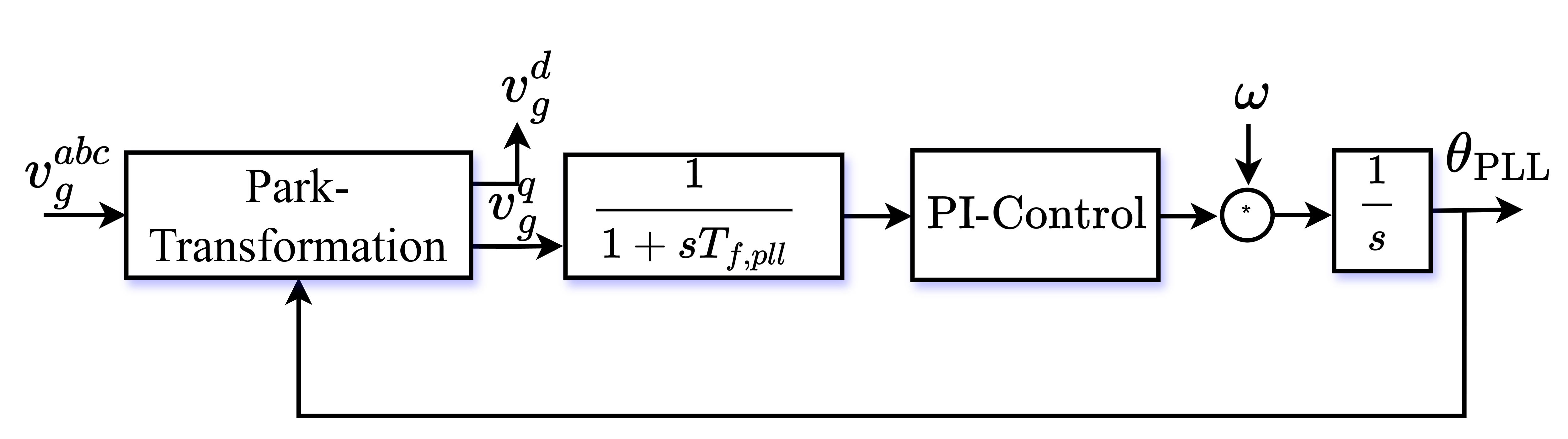} 
    \caption{PLL block diagram layout.}
    \label{fig4:PLL_diagram}
\end{figure}

The PLL operates by regulating the q-axis grid voltage component $v_g^q$ to zero using a PI controller. Forcing $v_g^q \rightarrow 0 $ ensures that the  d-axis aligns with the grid-voltage phasor, which in turn causes the $dq$ reference frame to rotate at the same angular speed as the grid voltage. The PI controller determines this rotation frequency, and integrating the controller output yields the estimated transformation angle used for the Park transformation. The general control block\footnote{The control equations are implemented using \cite{teodorescu2011grid} \cite{yazdani2010voltage}} diagram of PLL is represented in the Fig.\ref{fig4:PLL_diagram}

Before entering the PLL PI controller, the $q$-axis grid voltage $v_g^{q}(t)$ is passed through a first-order low-pass filter with time constant $T_{f,\mathrm{PLL}}$.  
This filter attenuates high-frequency noise in $v_g^{q}(t)$, In addition, the chosen value of $T_{f,\mathrm{PLL}}$ is used together with the symmetrical–optimum criterion\footnote{The method "symmetrically" places the crossover frequency of the open-loop system to maximize the phase margin, which in turn ensures the PLL is fast and also stable} to parameterize the PI gains of the PLL controller. The time constant of the filter provides a degree of freedom for tuning the controller gains. 

Let \( v_{g,fil}^{q} \) denote the filtered \( q \)-axis grid voltage obtained after passing 
\( v_{g,q} \) through a first-order low-pass filter. In the frequency domain, the filter is represented by
the transfer function
\begin{equation}\label{eq:pll_lpf}
H(s) = \frac{v_{g,fil}^{q}(s)}{v_{g}^{q}(s)} 
 = \frac{1}{1 + sT_{f,\mathrm{PLL}}}.
\end{equation}

The corresponding time-domain differential equation (equivalently) is
\begin{equation}\label{eq:pll_lpf_ode2}
\frac{d v_{g,fil}^{q}(t)}{dt}
=
\frac{1}{T_{f,\mathrm{PLL}}}
\left(
v_{g}^{q}(t)
-
v_{g,fil}^{q}(t)
\right).
\end{equation}


The next step is to pass it $v_{g,fil}^{q}$ to the PI controller with the reference set to zero.  
\begin{tcolorbox}[
    colback=gray!10,
    colframe=gray!50,
    title=PI controller,
    title style={color=black},
    enhanced,
]

The standard continuous-time PI controller is
\begin{equation}
u(t) = k_p\, e(t) + k_i \int_{0}^{t} e(\tau)\, d\tau,
\end{equation}
where $k_p$ is proportional gain, $k_i$ is integral gain, and $u(t)$ is the controller output. The basic PI controller can be written in the form of differential algebraic equations:
\begin{equation}
I(t) := k_i \int_{0}^{t} e(\tau)\, d\tau.
\end{equation}
Its dynamics are
\begin{equation}
\frac{d I(t)}{dt} = k_i\, e(t),
\end{equation}
and the controller output becomes
\begin{equation}
u(t) = k_p\, e(t) + I(t).
\end{equation}

\end{tcolorbox}

PI controller gains \(k_{p,\mathrm{PLL}}\) and \(k_{i,\mathrm{PLL}}\), and the internal PI state is \(I_{\mathrm{PLL}}(t)\). The controller output is denoted by \(\Delta_{\mathrm{PLL}}(t)\). The PI controller output $\Delta_{\mathrm{PLL}}(t)$ represents the per-unit frequency deviation. The corresponding physical frequency deviation in rad/s is obtained by multiplying it with $\omega$. 

\begin{align}
\frac{d I_{\mathrm{PLL}}(t)}{dt} &= k_{i,\mathrm{PLL}}\, v_{g,fil}^{q}(t),
\end{align}
and
\begin{equation}
\Delta_{\mathrm{PLL}}(t) = k_{p,\mathrm{PLL}}\, v_{g,fil}^{q}(t) + I_{\mathrm{PLL}}(t).
\end{equation}

\begin{equation}
\Delta \omega(t) = \omega \, \Delta_{\mathrm{PLL}}(t),    
\end{equation}

We compute the PLL angle deviation \( \theta_{\mathrm{PLL}}(t) \) by integrating the frequency deviation using the equivalent circuit shown in Fig.~\ref{fig:freq_integrator}.
\begin{equation}
\frac{d \theta_{\mathrm{PLL}}(t)}{dt}
= \Delta \omega(t)
= \omega \, \Delta_{\mathrm{PLL}}(t),
\qquad
\theta_{\mathrm{PLL}}(0) = 0.    
\end{equation}

\begin{figure}[H]
    \centering
    \includegraphics[width=0.9\textwidth]{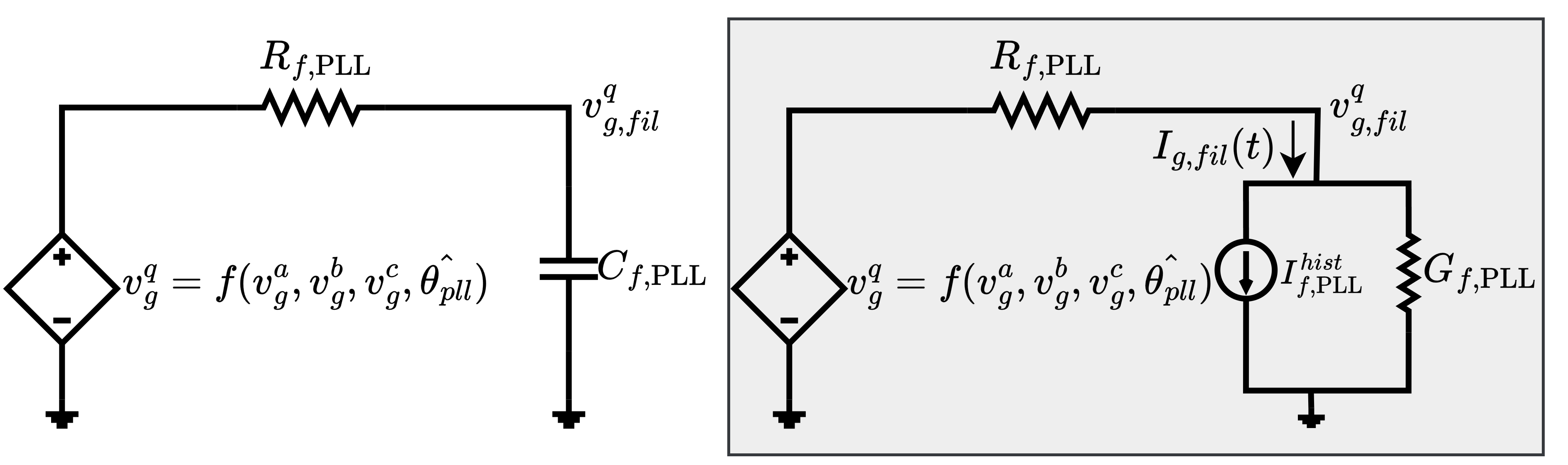} 
    \caption{Equivalent circuit implementation of the PLL low-pass filter using companion modeling.}
    \label{fig:PLL_LPF}
\end{figure}

We model \eqref{eq:pll_lpf_ode2} using an equivalent RC circuit Fig.~\ref{fig:PLL_LPF} and its companion model \cite{pillage1998electronic}, where
\( G_{f,\mathrm{PLL}} = \cfrac{2C_{f,\mathrm{PLL}}}{\Delta t} \) and
\( I_{f,\mathrm{PLL}}^{\mathrm{hist}}(t) = -G_{f,\mathrm{PLL}}\, v_{g,fil}^{q}(t-1) - I_{g,fil}(t-1) \).
The RC time constant is \( T_{f,\mathrm{PLL}} = R_{f,\mathrm{PLL}} C_{f,\mathrm{PLL}} \), and \( \Delta t \) denotes the trapezoidal discretization step.

Equivalently, we represent the PI controller and integrator stage of the PLL using their corresponding equivalent circuit models, as shown previously in Fig.~\ref{fig:PLL_LPF}.
\begin{figure}[H]
    \centering
    \includegraphics[width=0.7\textwidth]{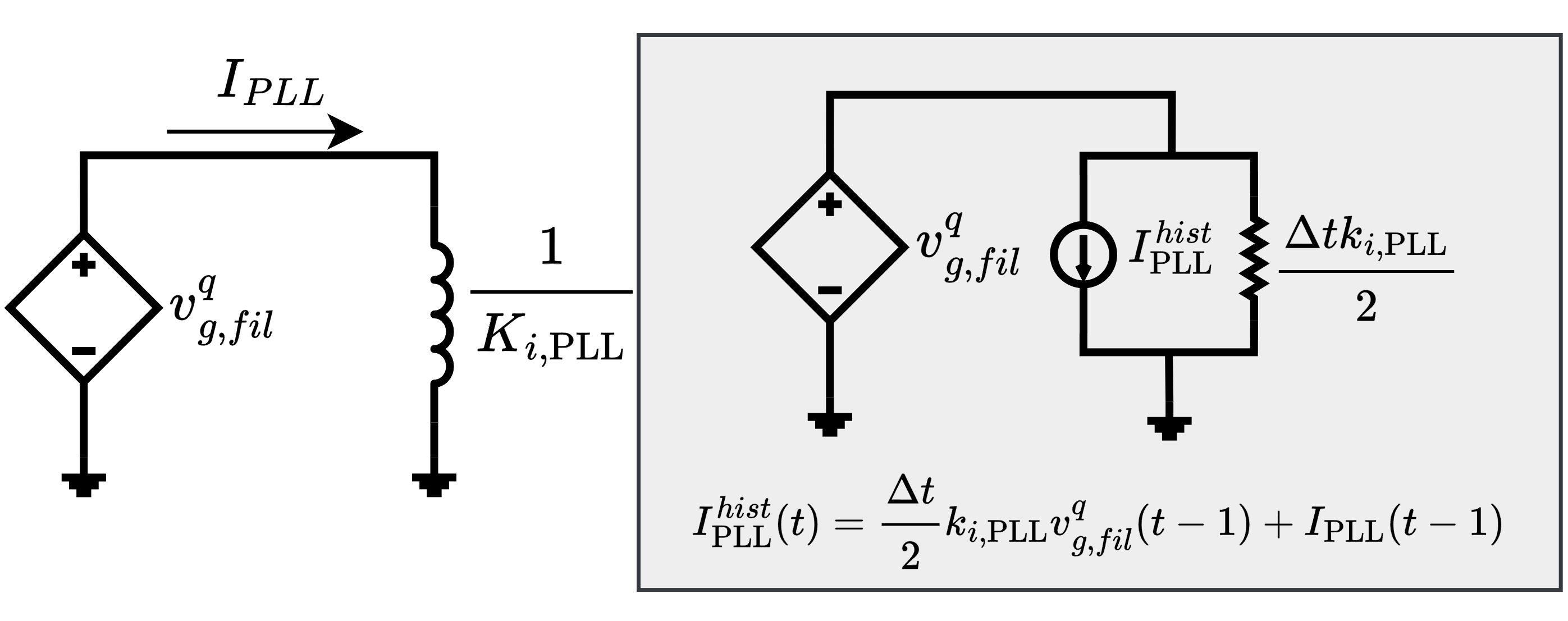} 
    \caption{Equivalent circuit implementation of the integral component of the PLL PI controller using companion modeling.}
    \label{fig:PLL_PI}
\end{figure}

\begin{figure}[H]
    \centering
    \includegraphics[width=0.7\textwidth]{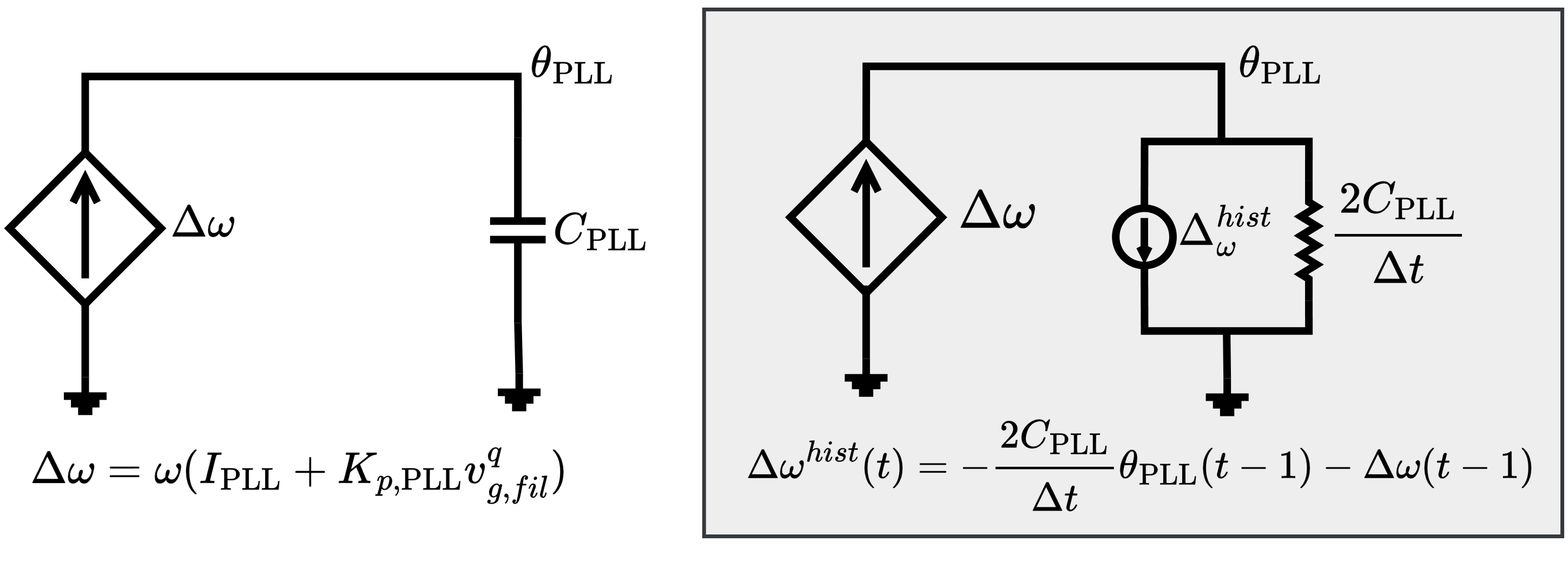} 
    \caption{Equivalent circuit implementation of the PLL integrator using companion modeling.}
    \label{fig:freq_integrator}
\end{figure}
\begin{figure}[H]
    \centering
    \includegraphics[width=0.95\textwidth]{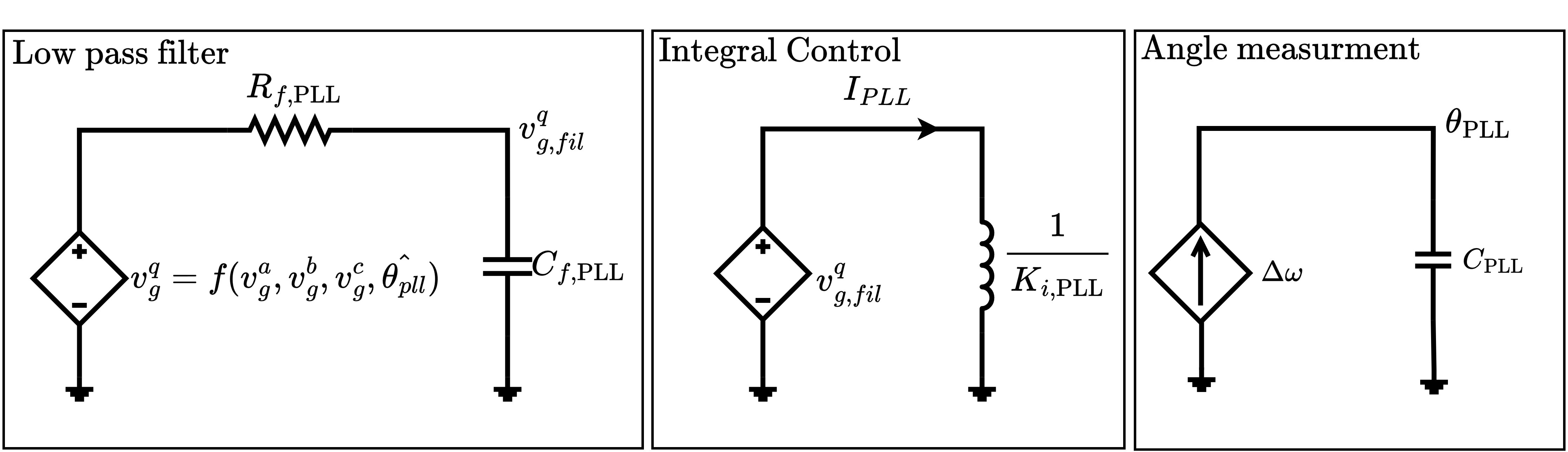} 
    \caption{Equivalent circuit model of PLL.}
    \label{fig:PLL_full}
\end{figure}

Building on the companion modeling framework introduced earlier, we assemble all PLL component circuits into a single equivalent diagram for implementation, as shown in Fig.~\ref{fig:PLL_full}. 


%% file: Section/Power_controller.tex
\section*{Power Controller (PC)}
The active and reactive power control is used as upper-level control and sets the current setpoints so that the desired (set) active and reactive power can be injected into the power system at the point of common coupling (PCC). The general power control blocks are represented in the following Fig.\ref{fig5:Power_controller_layout}. The power controller block has components of the current controller block which will be discussed in detail in the next section.  

\begin{figure}[H]
    \centering
    \includegraphics[width=0.8\textwidth]{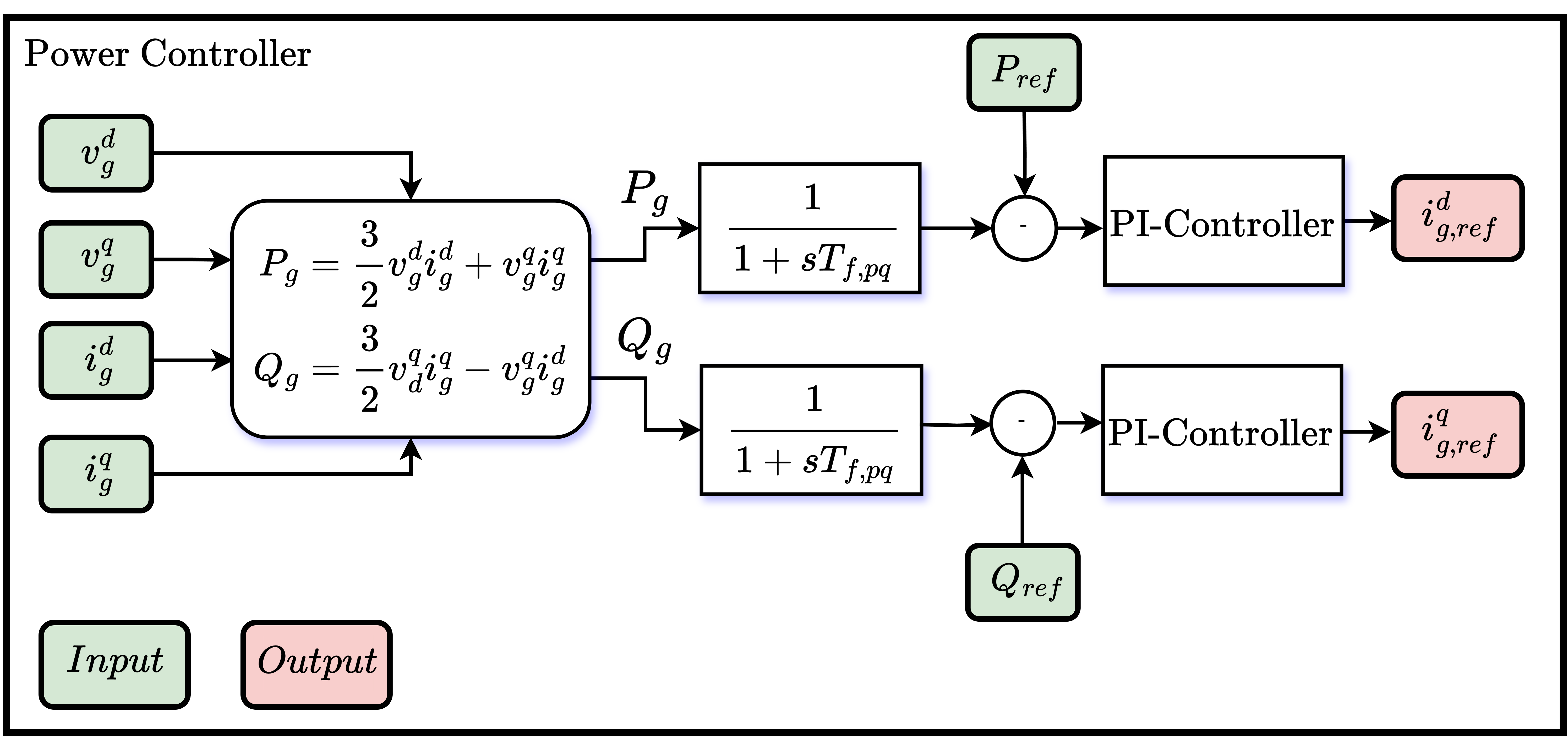} 
    \caption{Power controller blocks diagram layout.}
    \label{fig5:Power_controller_layout}
\end{figure}

The instantaneous active and reactive power at the PCC in the synchronous $dq$ frame are
computed as:
\begin{subequations}\label{eq:PQ_dq}
\begin{align}
    P_g(t) &= \frac{3}{2} \Big( v_g^{d}(t)\, i_g^{d}(t)
                              + v_g^{q}(t)\, i_g^{q}(t) \Big), \label{eq:P_dq} \\[2mm]
    Q_g(t) &= \frac{3}{2} \Big( v_g^{d}(t)\, i_g^{q}(t)
                              - v_g^{q}(t)\, i_g^{d}(t) \Big). \label{eq:Q_dq}
\end{align}
\end{subequations}

\begin{figure}[H]
    \centering
    \includegraphics[width=0.9\textwidth]{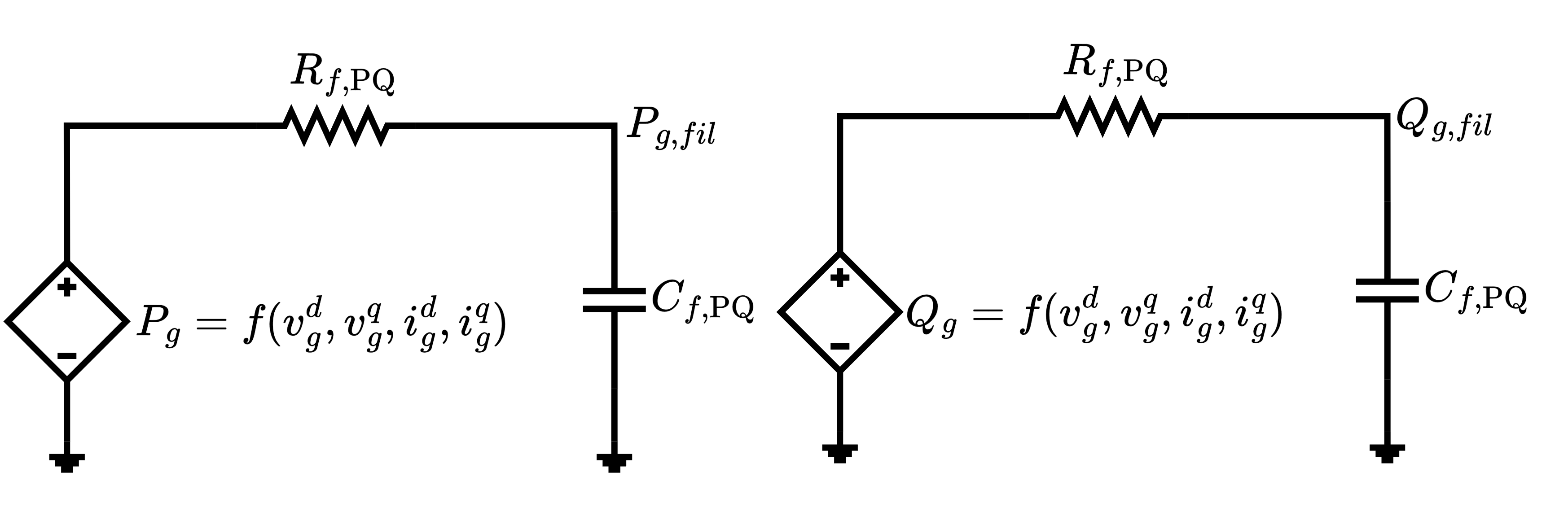} 
    \caption{Equivalent circuit model of the power-controller low-pass filter.}
    \label{fig:Power_control_LPF}
\end{figure}

The instantaneous power ($P_g(t)$, $Q_g(t)$)\eqref{eq:PQ_dq} is passed through the low-pass filter with time constant $T_{f,PQ} = (R_{f,PQ}) (C_{f,PQ})$, let $P_{g,fil}$ and $Q_{g,fil}$ be the filtered form of $P_g$ and $Q_g$ can be calculated by solving \eqref{eq:PQ_fil} or the equivalent circuit presented in Fig.\ref{fig:Power_control_LPF}. 

\begin{subequations}\label{eq:PQ_fil}
\begin{align}
\cfrac{d P_{g,fil (t)}}{dt} &= \cfrac{1}{T_{f,PQ}} \big(P_{g}(t) - P_{g,fil}(t) \big ) \\
\cfrac{d Q_{g,fil (t)}}{dt} &= \cfrac{1}{T_{f,PQ}} \big(Q_{g}(t) - Q_{g,fil}(t) \big )
\end{align}
\end{subequations}

\begin{figure}[H]
    \centering
    \includegraphics[width=0.8\textwidth]{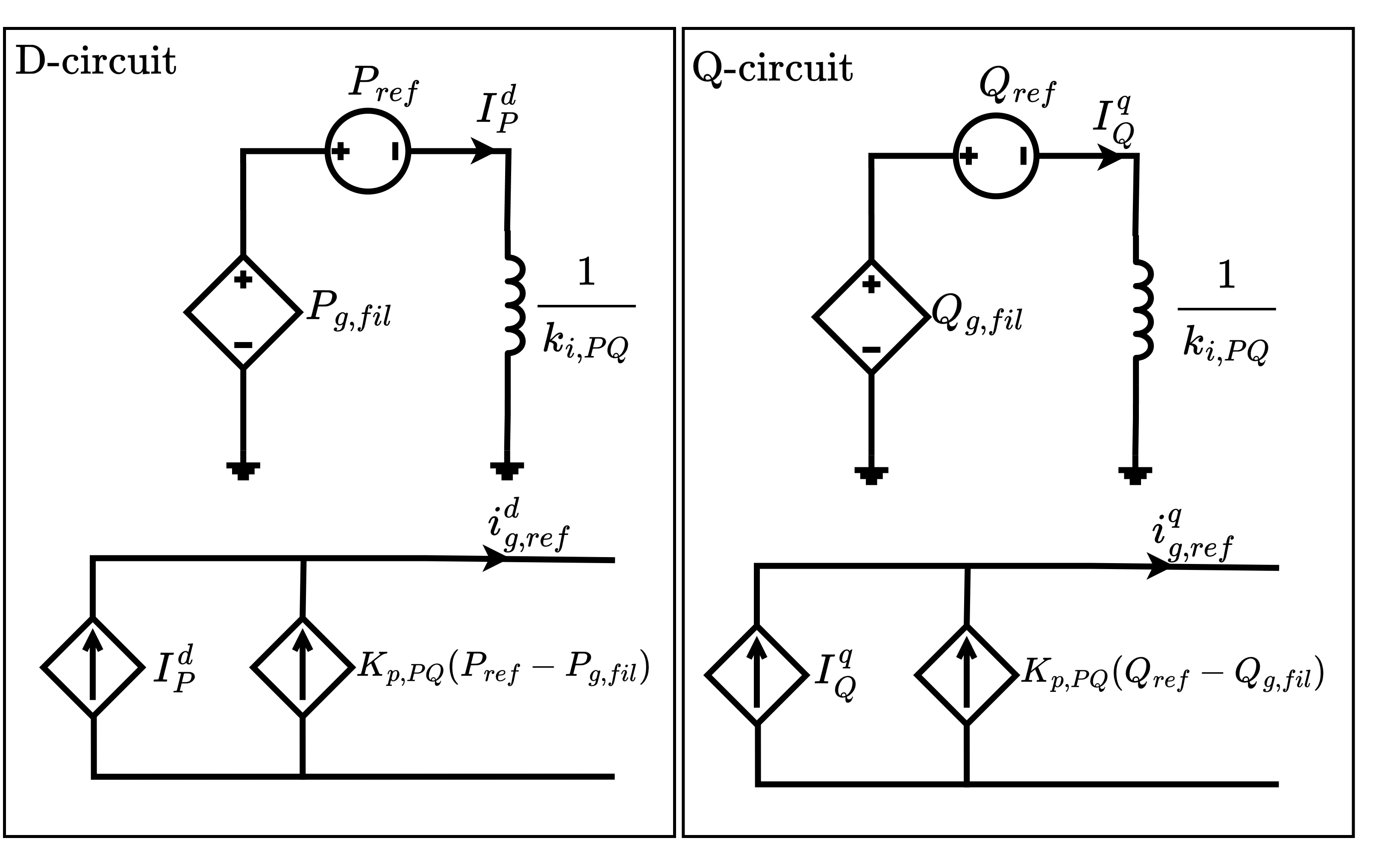} 
    \caption{Equivalent circuit representation of the power-controller PI controller.}
    \label{fig:Power_control_PI_controller}
\end{figure}

The filtered power is pass through the PI controller to generate the output $i_{g,ref}^d$ and $i_{g,ref}^q$ (\textit{later use in the current controller}); similar to the PLL equations, the controller proportional and integral gains are $k_{p,{PQ}}$ and $k_{i,{PQ}}$. 
\begin{subequations}\label{eq:ig_dref}
\begin{align}
\cfrac{d I_{P}^d(t)}{dt} = k_{i,\mathrm{PQ}} \big( P_{ref}(t) - P_{g,fil}(t) \big) \\
i_{g,ref}^{d} = k_{p,\mathrm{PQ}} \big( P_{ref}(t) - P_{g,fil}(t) \big) + I_{P}^{d}(t) \\
\cfrac{d I_{Q}^q(t)}{dt} = k_{i,\mathrm{PQ}} \big( Q_{ref}(t) - Q_{g,fil}(t) \big) \\
i_{g,ref}^{q} = k_{p,\mathrm{PQ}} \big( Q_{ref}(t) - Q_{g,fil}(t) \big) + I_{Q}^{q}(t) 
\end{align}
\end{subequations}
The equivalent circuit for both $D$ and $Q$ axis is shown in the Fig.\ref{fig:Power_control_PI_controller}. 

\begin{tcolorbox}[
    colback=gray!10,
    colframe=gray!50,
    title={Companion Model: Thevenin and Norton Equivalents},
    title style={color=black},
    enhanced,
]
    We can implement companion models using either their Norton-equivalent or Thevenin-equivalent circuit representations. Both formulations are mathematically equivalent and yield identical numerical solutions when solved consistently.

    The Norton-equivalent representation of a companion model is shown in Fig.~\ref{fig:companion_norton}. We solve this formulation by applying Kirchhoff’s Current Law (KCL) at the circuit nodes.

    \medskip
    \centering
    \includegraphics[width=0.4\linewidth]{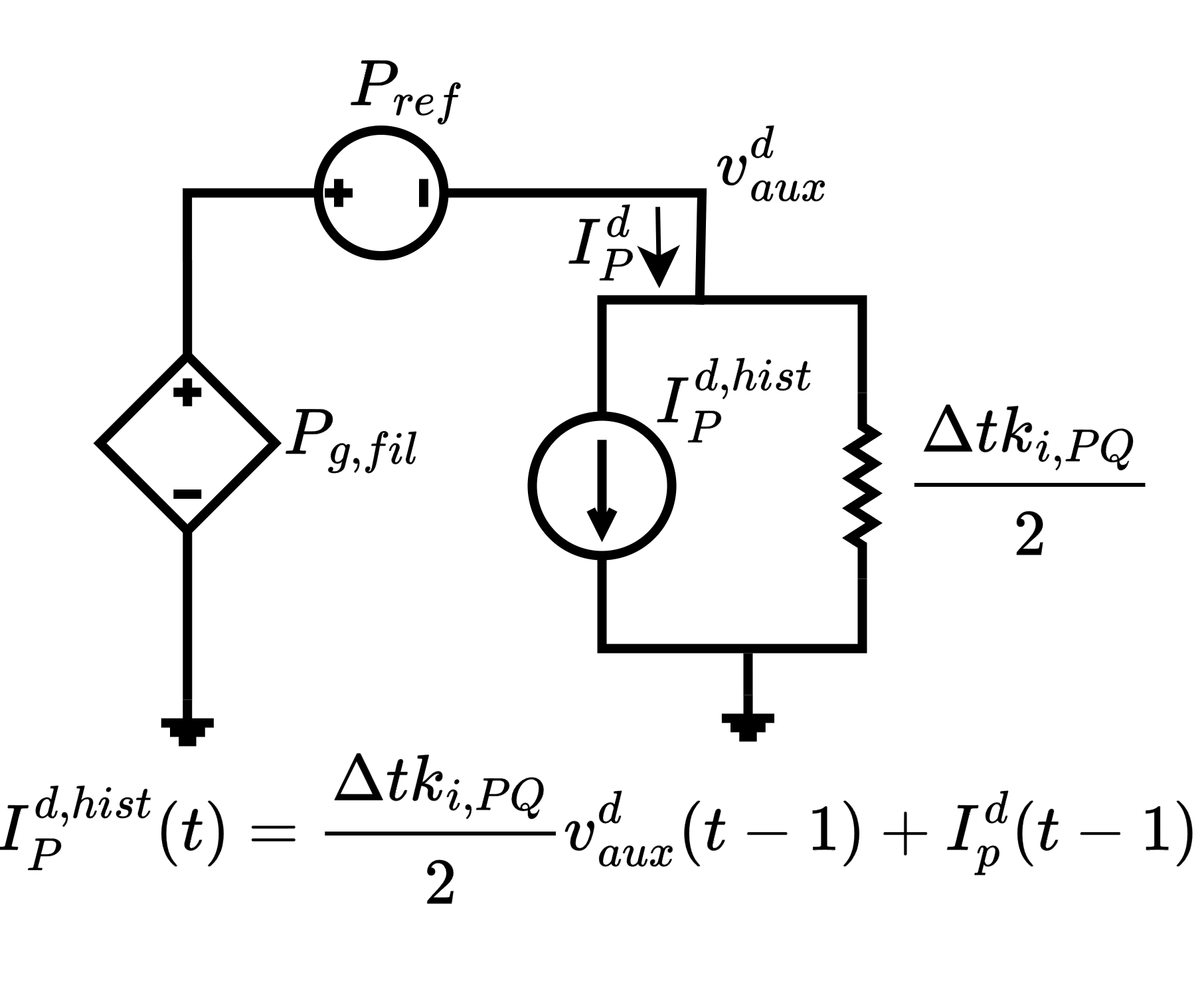}
    \captionof{figure}{Norton-equivalent companion model representation solved using KCL.}
    \label{fig:companion_norton}

    \medskip
    \raggedright 
    The corresponding Thevenin-equivalent representation of the same companion model is shown in Fig.~\ref{fig:companion_thevenin}. We solve this formulation by applying Kirchhoff’s Voltage Law (KVL) around the circuit loops.

    \medskip
    \centering
    \includegraphics[width=0.4\linewidth]{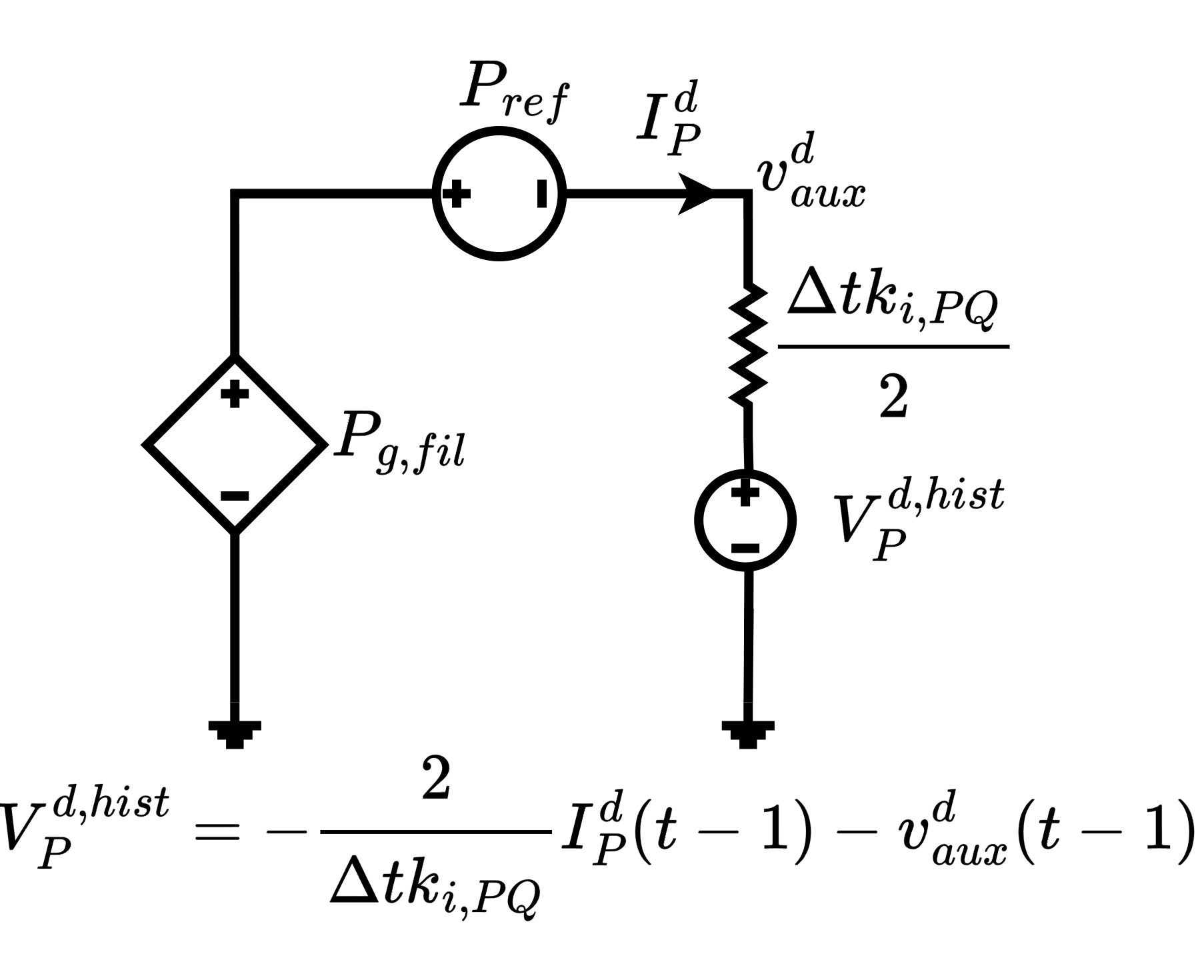}
    \captionof{figure}{Thevenin-equivalent companion model representation solved using KVL.}
    \label{fig:companion_thevenin}
    \medskip
    \raggedright
    Despite their different circuit forms, both representations describe the same discretized dynamic behavior. Consequently, solving either formulation produces identical state trajectories and network responses.
\end{tcolorbox}

%% file: Section/Current_controller_new.tex
\section*{Current Controller (CC)}
\begin{figure}[H]
    \centering
    \includegraphics[width=0.75\textwidth]{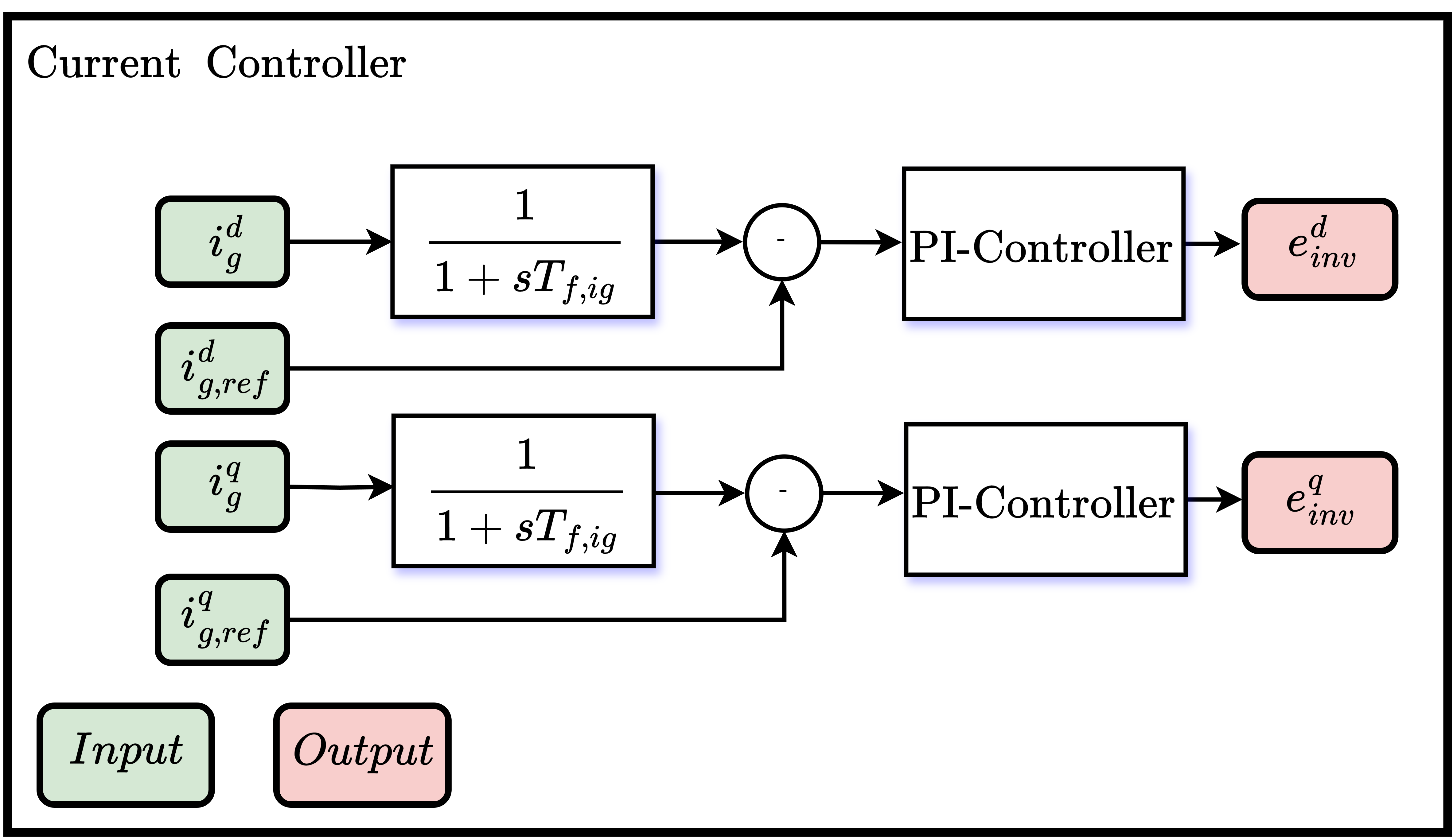} 
    \caption{Current controller block diagram layout.}
    \label{fig:current_controller_BD}
\end{figure}

Based on the grid-side dynamics, the current controller is used to control the grid currents $i_g^d$ and $i_g^q$, which results in the lower-level grid current controller. the general control block diagram of the current controller is presented in the Fig.\ref{fig:current_controller_BD}

\begin{figure}[H]
    \centering
    \includegraphics[width=0.7\textwidth]{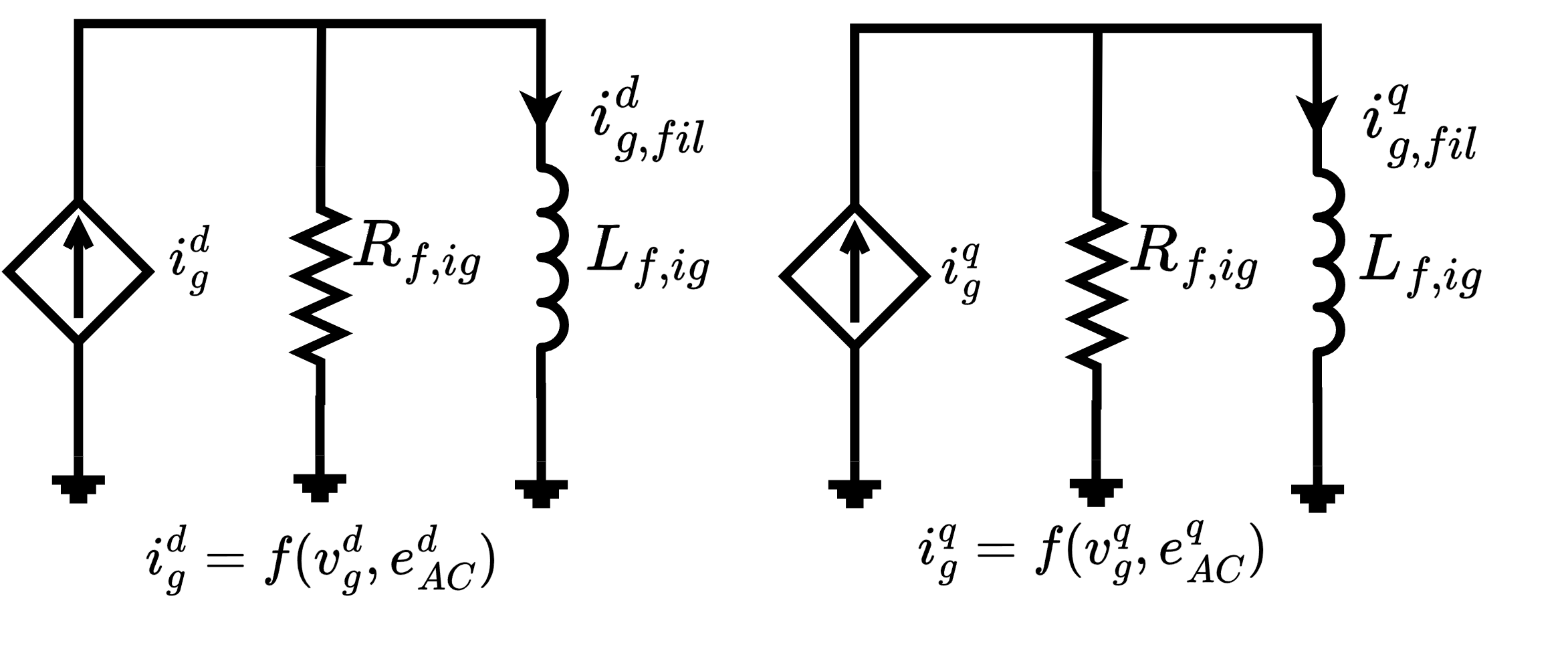} 
    \caption{Grid-Connected Inverter Equivalent Circuit Diagram.}
    \label{fig:current_controller_filter}
\end{figure}

The first part is to calculate grid currents $i_g^d$ and $i_g^q$ pass them through the low-pass filter with a time constant $T_{f,ig}$. 
The equivalent $i_g^d(t)$, $i_g^q(t)$ can be computed by applying the park's transformation to \eqref{eq:igdq_res} using the estimated PLL angle. 
Let $i_{g,fil}^d$ and $i_{g,fil}^q$ be the filtered currents used by the current controller. The equivalent circuits are represented in the Fig.\ref{fig:current_controller_filter}:

\begin{subequations}\label{eq:ig_fil}
\begin{align}
    \cfrac{d i_{g,fil}^d(t)}{dt} = \cfrac{1}{T_{f,ig}} \big(i_{g}^{d}(t) - i_{g,fil}^{d}(t) \big) \\
    \cfrac{d i_{g,fil}^q(t)}{dt} = \cfrac{1}{T_{f,ig}} \big(i_{g}^{q}(t) - i_{g,fil}^{q}(t) \big)
\end{align}    
\end{subequations}

\begin{figure}[H]
    \centering
    \includegraphics[width=0.7\textwidth]{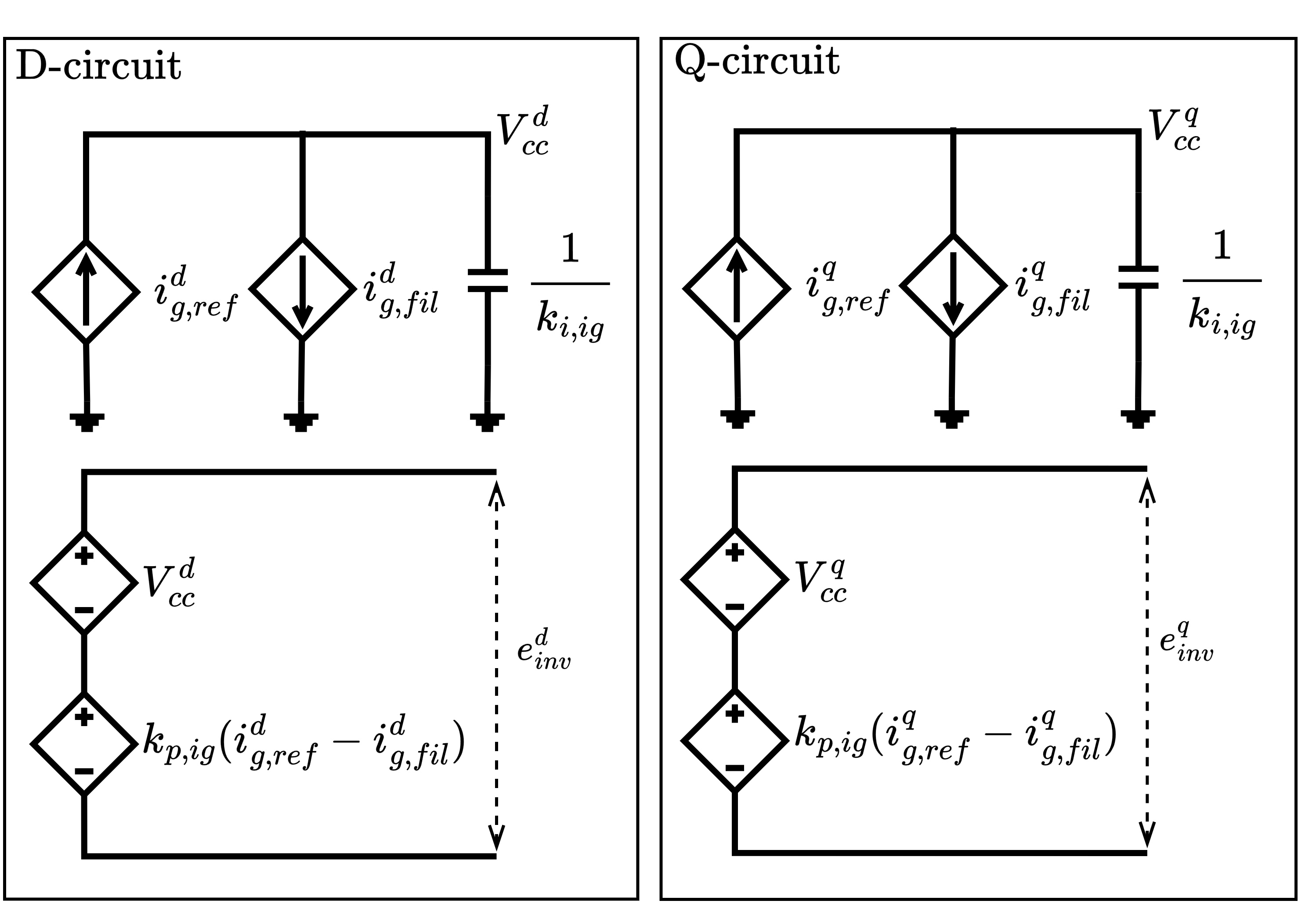} 
    \caption{Equivalent circuit representation of the current controller PI circuit.}
    \label{fig:current_controller_PI_controller}
\end{figure}

Now in a similar fashion, a PI controller is used to track the reference signal $i_{g,ref}^d$ $i_{g,ref}^q$ as implemented previously with $k_{p,\mathrm{ig}}$ and $k_{i,\mathrm{ig}}$ as the proportional and integral gains and can be implemented using the following circuit Fig.\ref{fig:current_controller_PI_controller}. 

\begin{subequations}\label{eq:PI_cc}
\begin{align}
    \cfrac{d V_{cc}^{d} (t)}{dt} = k_{i,\mathrm{ig}} \big(i_{g,ref}^d(t) - i_{g,fil}^d(t) \big) \\
    \epsilon^d(t) = k_{p,\mathrm{ig}} \big(i_{g,ref}^d(t) - i_{g,fil}^d(t) \big) + V_{cc}^{d} (t) \\
    \cfrac{d V_{cc}^q (t)}{dt} = k_{i,\mathrm{ig}} \big(i_{g,ref}^q(t) - i_{g,fil}^q(t) \big) \\
    \epsilon^q(t) = k_{p,\mathrm{ig}} \big(i_{g,ref}^q(t) - i_{g,fil}^q(t) \big) + V_{cc}^{q} (t)
\end{align}
\end{subequations}

If the filter inductor is not considered, then $e_{inv}^{d}(t) = \epsilon^d(t)$, and $e_{inv}^{q}(t) = \epsilon^q(t)$. 

\subsection*{Analysis including the filter inductor $L_f$}
For the analysis that includes the filter inductor dynamics, the grid phase voltages
$v_g^{a}$, $v_g^{b}$, and $v_g^{c}$ are explicitly modeled as state variables. The result of the PI controller is used with the system of equations, and $e_{inv}^{dq}$ equations can be derived as follows:
\begin{subequations}\label{eq:e_AC_dq}
\begin{align}
    e_{inv}^{d}(t) = \epsilon^d(t) - \hat{\omega}(t) L_f^{d} i_{g}^{q}(t) + v_g^d(t) \\
    e_{inv}^{q}(t) = \epsilon^q(t) + \hat{\omega}(t) L_f^{q} i_{g}^{d}(t) + v_g^q(t) 
\end{align}
\end{subequations}
In eq.\eqref{eq:e_AC_dq}, $\epsilon^d(t)$ and $\epsilon^q(t)$ are the error from the PI controller, and $\hat{\omega}(t)$ is the angular frequency and can be calculated from the PLL, $\Delta_{\mathrm{PLL}}(t)$, by using the formula:$\hat{\omega}(t)=\omega(1+\Delta_{\mathrm{PLL}}(t))$.  
\begin{tcolorbox}[
    colback=gray!10,
    colframe=gray!50,
    title=Feedforward Control/Disturbance Compensator,
    title style={color=black},
    enhanced
]
In addition to the PI controllers, disturbance compensation is performed by adding $v_g^d $, $v_g^q$ along with cross-coupling terms for the independent control of d- and q-currents. Mathematically (this analysis is for the d-axis; same can be applied to the q-axis):

\begin{equation}
\underbrace{(\text{PI-Controller} - \hat{\omega}(t) L_f^{d} i_g^q(t) + v_{g}^d)}_{\text{Controller Output } (e_{inv}^d)} = \underbrace{R_f i_g^d(t) + L_f^{d} \frac{di_g^d(t)}{dt} - \hat{\omega} L_f^{d} i_g^q(t) + v_{g}^d(t)}_{\text{Physical Plant Voltage}}
\end{equation}
Canceling the common terms on both sides yields:
\begin{equation}
    \text{PI-Controller} = R_f^{d} i_g^d(t) + L_f^{d} \frac{di_g^d(t)}{dt}
\end{equation}
Consequently, the PI controller sees the plant as a decoupled, linear first-order system determined solely by the filter resistance and inductance; this also helps in simplifying the controller gain tuning \cite{hahn2018modellierung}.
\end{tcolorbox}

The equivalent $i_g^d(t)$, $i_g^q(t)$ expressions in continuous time are
\begin{subequations}\label{eq:KVL_equations}
    \begin{align}
        L_f^{d}  \cfrac{di_g^d(t)}{dt} = -R_f^di_g^d(t) + \hat{\omega}(t) L_f^d i_g^q(t) + e_{inv}^d(t) - v_{g}^d(t) \\
        L_f^{d} \cfrac{di_g^q(t)}{dt} = -R_f^d i_g^q(t) - \hat{\omega}(t) L_f^d i_g^d(t) + e_{inv}^q(t) - v_{g}^q(t) 
    \end{align}
\end{subequations}

\begin{tcolorbox}[
    colback=gray!10,
    colframe=gray!50,
    title=Cross coupling terms,
    title style={color=black},
    enhanced
]
When differentiating in a dq-frame, an extra term appears as shown in \eqref{eq:KVL_equations} due to the rotation at the angular speed $\hat{\omega}(t)$, there is a $90^o$ rotational matrix that creates the cross-coupling:
\begin{equation}
    J = 
\begin{bmatrix}
0 & -1 \\
1 & 0
\end{bmatrix}.
\end{equation}

\end{tcolorbox}

\begin{figure}[H]
    \centering
    \includegraphics[width=0.6\textwidth]{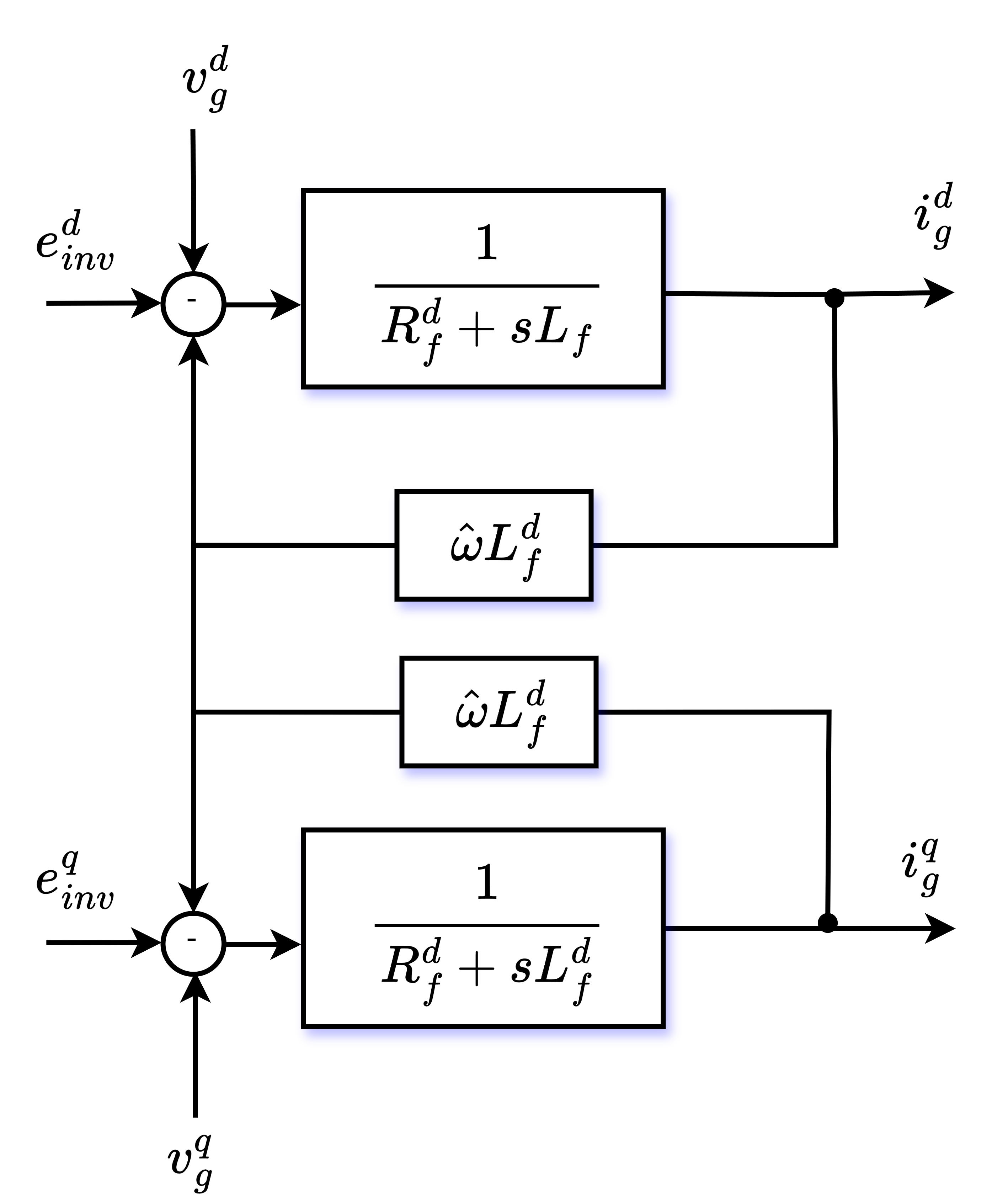} 
    \caption{$i_g^d(t)$, $i_g^q(t)$ Expressions Block Layout.}
    \label{fig7:id_iq_block_layout}
\end{figure}

The last part of the inverter modeling is the expression of $i_g^d(t)$, $i_g^q(t)$ which can be found by applying Kirchhoff's Voltage Law (KVL), Fig.\ref{fig7:id_iq_block_layout} represents the KVL applied to the Fig.\ref{fig1:inverter_diagram} circuit in the frequency domain. 

\noindent\textbf{Note:} Under balanced conditions where $R_f^a = R_f^b = R_f^c$, the resistances $R_f^d = R_f^q$, and the transformation from the phase domain $(abc)$ to the $dq0$ frame follows $R_f^{dq0} = \mathbf{P}R_f^{abc}\mathbf{P}^{-1}$, where, $\mathbf{P}$ is the Park's transformation matrix. The same principle applies to the filter inductances: balanced phase inductances yield constant and equal $dq$frame values, i.e., $L_f^d = L_f^q$. Accordingly, the values of $R_f^{abc}$, and $L_f^{abc}$ are provided in the code base.

%% file: Section/Grid_support.tex
\section*{Grid Support Tasks}
We extend the grid-following inverter model beyond basic power injection to include services that support grid stability. These services enable the inverter to respond dynamically to grid frequency and voltage deviations and thereby emulate key supportive behaviors traditionally provided by synchronous machines.

\subsection*{Frequency Support}
Frequency support, also called frequency-watt or droop control, allows the inverter to adjust its active power output in response to frequency fluctuations. We implement this functionality by modifying the active power reference according to the deviation of the estimated angular frequency, $\hat{\omega}(t)$, from the nominal frequency, $\omega$. To prevent unnecessary control action in response to small measurement noise or minor frequency variations, we also introduce a frequency deadband, $f_{db}$.

\begin{figure}[H]
    \centering
    \includegraphics[width=0.7\textwidth]{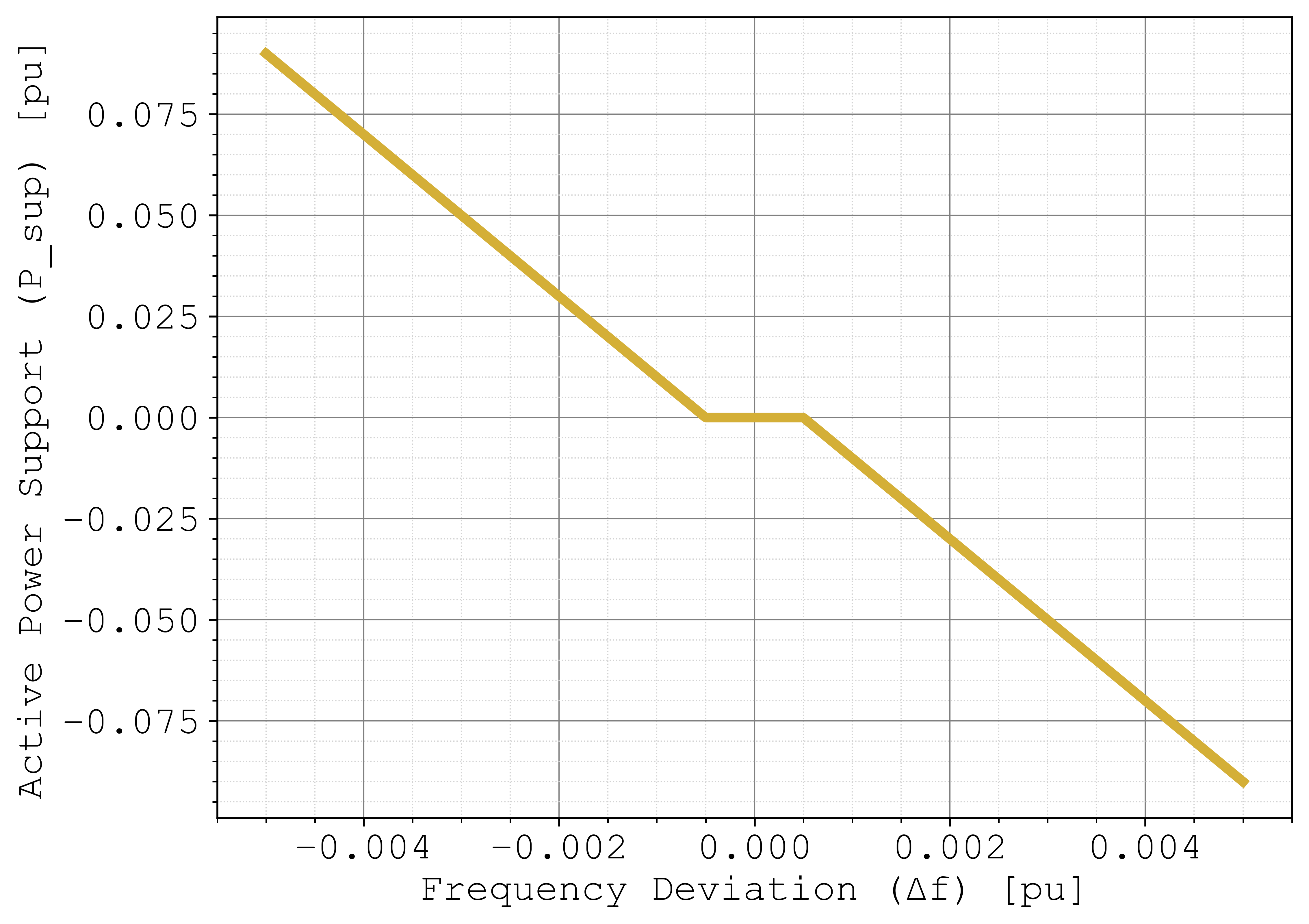} 
    \caption{Frequency-watt characteristic for active power support. The slope outside the deadband is set by the selected droop parameters.}
    \label{fig8:FS_curve}
\end{figure}

Figure~\ref{fig8:FS_curve} shows the frequency-watt droop characteristic adopted in this tutorial. We compute the supplementary active-power command, $P_{sup}$, from the frequency deviation, $\Delta f$, using a piecewise-linear law with a central deadband. This deadband prevents the controller from reacting to small frequency variations and measurement noise near nominal conditions. When $\Delta f$ exceeds the deadband limits, the inverter increases or decreases its active-power reference according to the prescribed droop slope. Therefore, the exact shape of the curve depends on the selected droop gain, deadband width, and any support-power limits imposed in the controller.

We augment the baseline active-power reference, $P_{ref}(t)$, with the supplementary frequency-support signal, $P_{sup}(t)$, to obtain the total active-power reference, $P_{ref,total}(t)$.

\begin{equation}
    P_{ref}(t)\;\mapsto\;P_{ref,total}(t) = P_{ref}(t) + P_{sup}(t)
\end{equation}
The equations represent the Fig.~\ref{fig8:FS_curve}, where $k_f$ is the droop co-efficient. 
\begin{subequations}\label{eq:frequency_support}
\begin{align}
\Delta f(t) &= \frac{\hat{\omega}(t) - \omega}{\omega}
\label{eq:frequency_deviation}
\\
P_{\mathrm{sup}}(t) &=
\begin{cases}
-k_f\left(\Delta f(t)-f_{db}\right), & \text{if } \Delta f(t) > f_{db}, \\[0.5ex]
-k_f\left(\Delta f(t)+f_{db}\right), & \text{if } \Delta f(t) < -f_{db}, \\[0.5ex]
0, & \text{otherwise,}
\end{cases}
\label{eq:psup}
\end{align}
\end{subequations}

\subsection*{Volt-Var Support}
Volt-Var support allows the inverter to regulate the voltage at the point of common coupling (PCC) by adjusting its reactive power output. We implement this control by monitoring the voltage magnitude at the PCC and modifying the reactive-power reference according to the deviation from a target voltage. To prevent unnecessary control action near nominal operating conditions, we introduce a voltage deadband, $V_{db}$. We also limit the supplementary reactive-power command to $Q_{\max,\mathrm{sup}}$ so that the inverter remains within its reactive-power capability.

We augment the baseline reactive-power reference, $Q_{ref}(t)$, with the supplementary Volt-Var support signal, $Q_{sup}(t)$, to obtain the total reactive-power reference, $Q_{ref,total}(t)$.

\begin{figure}[H]
    \centering
    \includegraphics[width=0.7\textwidth]{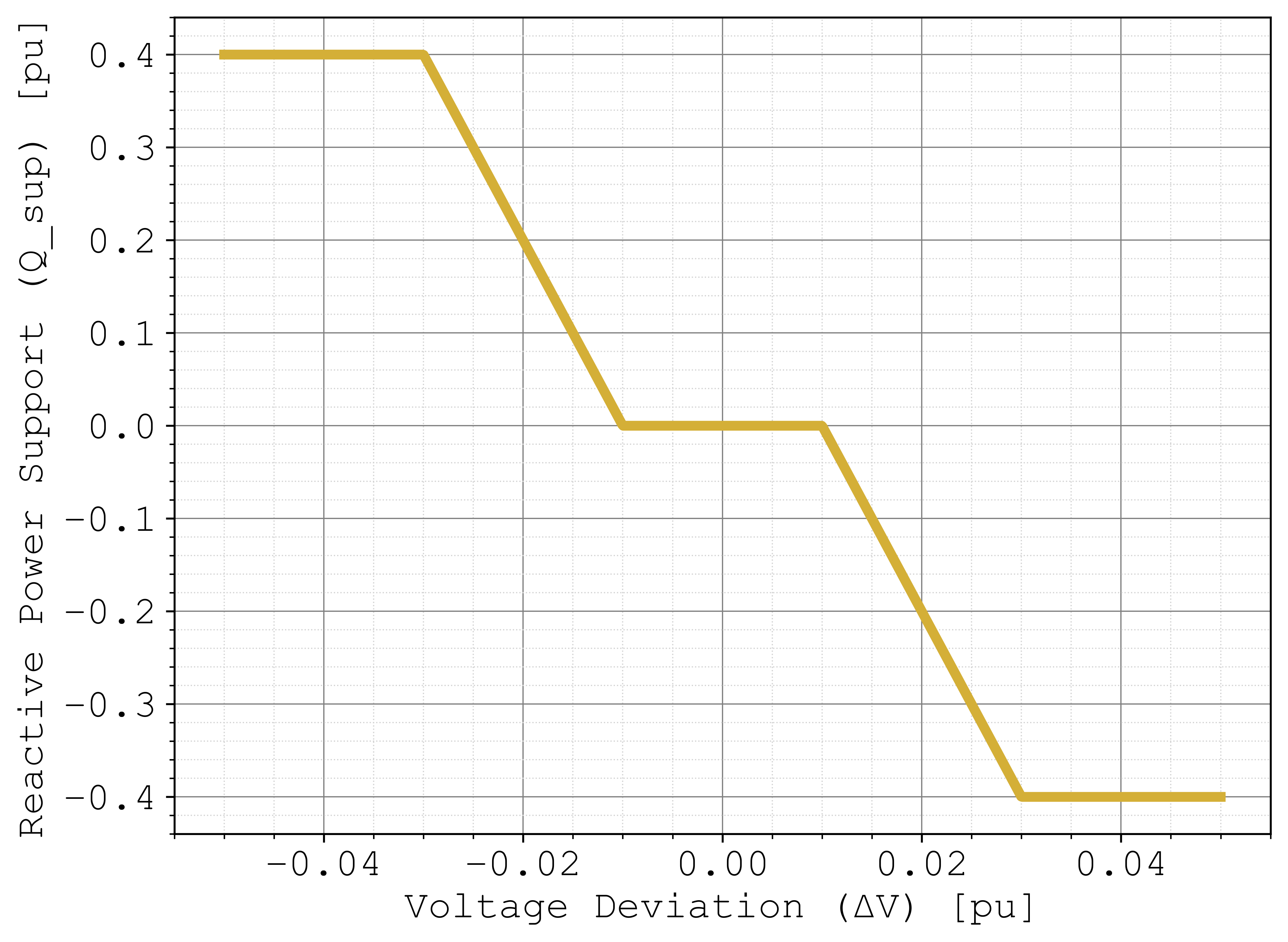} 
    \caption{Volt-Var characteristic for reactive power support. The slope outside the deadband is set by the selected droop parameters.}
    \label{fig9:VV_curve}
\end{figure}

\begin{equation}
    Q_{ref}(t)\;\mapsto\;Q_{\mathrm{ref,total}}(t) = Q_{\mathrm{ref}}(t) + Q_{\mathrm{sup}}(t)
\end{equation}
The equations represent the Fig.~\ref{fig9:VV_curve}, where $k_v$ is the droop co-efficient. 
\begin{subequations}\label{eq:volt_var_support}
\begin{align}
\Delta V(t) &= v_g^{d}(t) - V_{\mathrm{target}}
\label{eq:voltage_deviation}
\\
Q_{\mathrm{val}}(t) &=
\begin{cases}
-k_v\left(\Delta V(t)-V_{db}\right), & \text{if } \Delta V(t) > V_{db}, \\[0.5ex]
-k_v\left(\Delta V(t)+V_{db}\right), & \text{if } \Delta V(t) < -V_{db}, \\[0.5ex]
0, & \text{otherwise,}
\end{cases}
\label{eq:qval}
\\
Q_{\mathrm{sup}}(t) &= \max\!\left(-Q_{\max,\mathrm{sup}},\,
\min\!\left(Q_{\max,\mathrm{sup}},\,Q_{\mathrm{val}}(t)\right)\right)
\label{eq:qsup}
\end{align}
\end{subequations}

\subsection*{Handling non-differentiability}
The non-differentiable if-else deadband logic can hinder Newton-Raphson convergence by introducing Jacobian discontinuities. To address this issue, we replace it with smooth algebraic approximations \cite{badmus2026two}.

\begin{subequations}\label{eq:smooth_max_min}
\begin{align}
\mathrm{smax}(a,b,\epsilon) &= \frac{a+b+\sqrt{(a-b)^2+\epsilon}}{2}
\label{eq:smax}
\\
\mathrm{smin}(a,b,\epsilon) &= \frac{a+b-\sqrt{(a-b)^2+\epsilon}}{2}
\label{eq:smin}
\end{align}
\end{subequations}
where $\epsilon$ is a small positive smoothing parameter, for example $\epsilon = 10^{-5}$. We use these smooth operators to construct differentiable expressions for the support signals $P_{sup}$ and $Q_{sup}$ across the deadband boundaries. 

We use \eqref{eq:smooth_max_min} to construct smooth approximations of the non-differentiable characteristics shown in Fig.~\ref{fig8:FS_curve} and Fig.~\ref{fig9:VV_curve}.

\begin{subequations}\label{eq:smooth_support_functions}
\begin{align}
P_{\mathrm{sup}}(t) &= \mathrm{smin}\!\big(0,\,-k_f\big(\Delta f(t)-f_{db}\big),\,\epsilon\big)
+ \mathrm{smax}\!\big(0,\,-k_f\big(\Delta f(t)+f_{db}\big),\,\epsilon\big)
\label{eq:psup_smooth}
\\
Q_{\mathrm{val}}(t) &= \mathrm{smin}\!\big(0,\,-k_v\big(\Delta V(t)-V_{db}\big),\,\epsilon\big)
+ \mathrm{smax}\!\big(0,\,-k_v\big(\Delta V(t)+V_{db}\big),\,\epsilon\big)
\label{eq:qval_smooth}
\\
Q_{\mathrm{temp}}(t) &= \mathrm{smin}\!\big(Q_{\mathrm{val}}(t),\,Q_{\max,\mathrm{sup}},\,\epsilon\big)
\label{eq:qtemp_smooth}
\\
Q_{\mathrm{sup}}(t) &= \mathrm{smax}\!\big(Q_{\mathrm{temp}}(t),\,-Q_{\max,\mathrm{sup}},\,\epsilon\big)
\label{eq:qsup_smooth}
\end{align}
\end{subequations}

We include the full implementation used in this tutorial in the accompanying code repository at \href{https://github.com/hamzaali412/Tutorial-Gird-Following-Inverter.git}{\textit{GitHub repository}}, so readers can reproduce the reported results.

%% file: Section/Initial_conditions.tex
\section*{Appendix}
For the initialization, the brief discussion is provided in this  \href{https://arxiv.org/abs/2212.12368}{document}\cite{pandey2023circuit}, the following table provides the initial conditions that can be used for solving the circuit at $t=0$.
\begin{table}[h!]
    \centering
    \renewcommand{\arraystretch}{1.5}

    \caption{Initialization at $t=0$}
    \label{tab:initial_conditions}

    \begin{tabular}{@{} l l @{}} 
        \toprule
        \textbf{Variable} & \textbf{Initial Condition} \\ 
        \midrule
        $v_{g,fil}^q$ & $0$ \\ 
        $I_{PLL}$ & $0$ \\
        $\theta_{PLL}$ & $\theta_{grid}$ \\
        $P_{g,fil}$  & $P_g(0)$ \\
        $Q_{g,fil}$  & $Q_g(0)$ \\
        $I_{P}^d$ & $\cfrac{2}{3}\cfrac{P_g(0)}{V_m}$ \\
        $I_{Q}^q$ & $\cfrac{2}{3}\cfrac{Q_g(0)}{V_m}$ \\
        $i_{g,fil}^d$ & $\cfrac{2}{3}\cfrac{P_g(0)}{V_m}$ \\
        $i_{g,fil}^q$ & $\cfrac{2}{3}\cfrac{Q_g(0)}{V_m}$ \\
        $I_{CC}^d$ & $R_g \left(\cfrac{2}{3}\cfrac{P_g(0)}{V_m}\right)$ \\
        $I_{CC}^q$ & $R_g \left(\cfrac{2}{3}\cfrac{Q_g(0)}{V_m}\right)$ \\
        $i_{g}^d$ & $\cfrac{2}{3}\cfrac{P_g(0)}{V_m}$ \\
        $i_{g}^q$ & $\cfrac{2}{3}\cfrac{Q_g(0)}{V_m}$ \\
        \bottomrule
    \end{tabular}
\end{table}